\documentclass[fleqn,usenatbib]{mnras}

\usepackage{newtxtext,newtxmath}
\usepackage[T1]{fontenc}
\usepackage{ae,aecompl}

\usepackage{graphicx}	% Including figure files
\usepackage{amsmath}	% Advanced maths commands
\usepackage{multicol}

\usepackage{fontawesome5}
\usepackage{hyperref}

\makeatletter
\newcommand{\github}[1]{%
   \href{#1}{\faGithubSquare}%
}
\makeatother

\newcommand{\correctionsbold}{%
}

\newcommand{\microns}{\,$\upmu\mathrm{m}$} 

\title[Detecting gravitational lenses using ML]{Detecting gravitational lenses using machine learning: exploring interpretability and sensitivity to rare lensing configurations}

\author[J. Wilde et al.]{
Joshua Wilde,$^{1}$\thanks{E-mail: joshua.wilde@open.ac.uk}
Stephen Serjeant,$^{1}$
Jane M. Bromley,$^{2}$
Hugh Dickinson$^{1}$
\newauthor L\'eon V. E. Koopmans$^{3}$
R. Benton Metcalf$^{4,5}$
\\
$^{1}$School of Physical Sciences, The Open University, Walton Hall, Milton Keynes, MK7 6AA, UK\\
$^{2}$School of Computing \& Communications, The Open University, Walton Hall, Milton Keynes, MK7 6AA, UK\\
$^{3}$Kapteyn Astronomical Institute, University of Groningen, PO Box 800, 9700AV Groningen, The Netherlands\\
$^{4}$Dipartimento di Fisica e Astronomia, Università di Bologna, via Gobetti 93/2, I-40129 Bologna, Italy\\
$^{5}$INAF - Osservatorio di Astrofisica e Scienza dello Spazio di Bologna, via Gobetti 93/3, I-40129 Bologna, Italy
}

\date{Accepted XXX. Received YYY; in original form ZZZ}

\pubyear{2021}

\begin{document}
\label{firstpage}
\pagerange{\pageref{firstpage}--\pageref{lastpage}}
\maketitle

%\newcommand{\bfreferee}[0]{}
%\newcommand{\bfreferee}[0]{\bf}

% Abstract of the paper
\begin{abstract}
Forthcoming large imaging surveys such as Euclid and the Vera Rubin Observatory Legacy Survey of Space and Time are expected to find more than $10^5$ strong gravitational lens systems, including many rare and exotic populations such as compound lenses, but these $10^5$ systems will be interspersed among much larger catalogues of $\sim10^9$ galaxies. This volume of data is too much for visual inspection by volunteers alone to be feasible and gravitational lenses will only appear in a small fraction of these data which could cause a large amount of false positives. Machine learning is the obvious alternative but the algorithms' internal workings are not obviously interpretable, so their selection functions are opaque and it is not clear whether they would select against important rare populations. We design, build, and train several Convolutional Neural Networks (CNNs) to identify strong gravitational lenses using VIS, Y, J, and H bands of simulated data, with F1 scores between 0.83 and 0.91 on 100,000 test set images. We demonstrate for the first time that such CNNs do not select against compound lenses, obtaining recall scores as high as 76\% for compound arcs and 52\% for double rings. We verify this performance using Hubble Space Telescope (HST) and Hyper Suprime-Cam (HSC) data of all known compound lens systems. Finally, we explore for the first time the interpretability of these CNNs using Deep Dream, Guided Grad-CAM, and by exploring the kernels of the convolutional layers, to illuminate why CNNs succeed in compound lens selection.
\setcounter{footnote}{0}
\renewcommand*{\thefootnote}{\arabic{footnote}} 
\footnotemark

\end{abstract}

% Select between one and six entries from the list of approved keywords.
% Don't make up new ones.

%%%%%%%%%%%%%%%%%%%%%%%%%%%%%%%%%%%%%%%%%%%%%%%%%%

%%%%%%%%%%%%%%%%% BODY OF PAPER %%%%%%%%%%%%%%%%%%

\section{Introduction}
In the near future the Euclid Space Telescope \citep{laureijs2011euclid} and the Vera Rubin Observatory Legacy Survey of Space and Time (LSST) \citep{abell2009lsst} will achieve first light. Both Euclid and Rubin will be mapping about half the sky in optical/near-infrared bands to AB depths of $\sim24-24.5$ (Euclid) and $\sim22-25$ (LSST). Euclid is expected to find more than $10^5$ strong gravitational lens systems \citep[e.g.][]{collett2015} and 5000 strongly lensed quasars while LSST expects to find $10^5$ galaxy-galaxy lenses, $10^4$ strongly lensed quasars and 500 lensed type Ia supernovae with 100 of these lenses being suitable for gathering time-delay data \citep{zhan2018cosmology,verma2019strong}.
\setcounter{footnote}{1}
\footnotetext{
The code used in this paper is publicly available at \url{https://github.com/JoshWilde/LensFindery-McLensFinderFace} \github{https://github.com/JoshWilde/LensFindery-McLensFinderFace.
}}

With this large number of gravitational lenses, several cosmologically important quantities can be constrained. These include the dark energy equation of state (\correctionsbold{$w$}) parameter, the Hubble constant ($H_0$), and dark matter halo substructure. The current tension in $H_0$ could be eased (or at least illuminated) by measuring the time delays caused by gravitational lensing. Currently $H_0$ can only be constrained through lensing to  a precision of \correctionsbold{2.4}\% using this method (including the error budget for systematics and assuming a spatially flat cosmology) due to the limited number of suitable gravitational lenses that are currently known 
\correctionsbold{\citep{wong2020h0licow}. The problem with using time delays from gravitational lenses is that the assumption of the mass distribution using methods such as the mass-sheet transformation can make all observable strong lensing parameters invariant apart from the time delay \citep{schneider2014source}. The time delay of a lens is roughly $\Delta t \propto H_0^{-1}(1-k_e)$, where $k_e$ is the mean convergence. The only gravitational lens observable which directly constrains this value is the time delay. Properties such as stellar kinematics can provide additional constraints on $k_e$. Overconstrained models can generate a value for $k_e$ that are precise but not accurate resulting in uncertainties that are far larger than reported \citep{kochanek2020overconstrained}. It is possible to break this degeneracy without assuming a mass profile, instead adding spatially resolved kinematics from time delay and non-time delay lens systems resulting in a precision of 2.5\% \citep{birrer2020tdcosmo}} The amount of data from Euclid and LSST will increase the number of suitable gravitational lenses for measuring time delays. This is expected to allow for a determination of $H_0$ with an accuracy of less than 1\% \citep{liao2019hubble}.

Such a large sample of strong gravitational lenses should also be an excellent source of rare and exotic lens configurations. Compound lenses (multiple lens plane systems) are both rare and extremely valuable to cosmology. Over the course of its mission LSST expects to find $\sim90$ compound lens systems \citep{mandelbaum2018lsst}. These systems enforce tighter constraints on models allowing for lower uncertainties in measurements of \correctionsbold{parameters} such as $\Omega_M$ and \correctionsbold{$w$} \citep{collett2012constraining, collett2014cosmological}. From as few as 50 compound lenses these values can be measured to within a 10\% accuracy \citep{gavazzi2008sloan}. \correctionsbold{Other parameters such as $w$ and $\Omega_m$ can be determined with a high accuracy from the time delays of 100 lensed quasars \citep{treu2013dark}.}

However, Euclid's entire non-lensed catalogue will comprise about a billion galaxies, and LSST's will be even larger. Finding strong gravitational lenses, and exotic compound systems in particular, will therefore be a ``needle in a haystack in a field of haystacks'' data mining problem. In this paper we define compound systems to be a gravitational lens containing \correctionsbold{two sets of strongly lensed images}. The imaging data are too numerous to rely on human inspection alone \citep[e.g.][]{marshall_spacewarps_2016,more2016space,sonnenfeld_spacewarps_2020}, but machine learning is a natural approach to attempt to solve this problem. 

Machine learning was originally inspired by the processes of neurons within the brain and was initially developed as artificial neural networks \citep{mcculloch1943logical}. These could only perform simple tasks such as playing checkers \citep{samuel1960programming}. These artificial neural networks were further improved upon with algorithms such as backpropagation \citep{rumelhart1986learning} allowing for information to be transferred between hidden layers. Taking inspiration from how the human eye processes information, Convolutional Neural Networks (CNNs) were created \citep{lecun1989backpropagation,lecun1998gradient}. These included convolutional layers at the start of the network that forced the CNNs to learn information about how features were related within the image. These layers tend to start off by learning simplistic features such as edges, then they progress into patterns and shapes, eventually detecting complex features within the data. 

Over the past decade, machine learning and in particular CNNs have become more prominent following from advances in GPUs, the creation of large accessible datasets \citep{deng2009imagenet}, and the success of the deep learning CNN AlexNet \citep{krizhevsky2012imagenet}. As it has become popular in the mainstream, machine learning has been increasingly used in astronomy, in particular to classification problems, such as: using CNNs and multi-band images to classify dwarf galaxies \citep{muller2021dwarfs}, using CNNs to assign Fanaroff-Riley classifications to radio galaxies  \citep{scaife2021fanaroff}, using an autoencoder for morphological classification of galaxies \citep{spindler2021astrovader}, using a U-Net to perform source detection, segmentation, and classification  \citep{hausen2020morpheus}, using RNN to correct classifications in maps \citep{maggiori2017recurrent}, and classifying galaxies using T-SNE  \citep{zhang2020powerful}. Denoising autoencoders have also been used for image deconvolution \citep{lauritsen2021super}.

In recent years machine learning has been applied to search for gravitational lenses.  
Various machine learning techniques have been used for this application, which can be broken down into two main categories: supervised and unsupervised learning.
Supervised machine learning approaches require large amounts of data in order to sensibly train. Since the number of lenses discovered is much smaller than the number of lenses need to train machine learning approaches, simulated data is often used for this task \citep[e.g.][]{khramtsov2019kids,he2020deep,jacobs2017finding,huang2019finding,jacobs2019finding,hezaveh2017fast, petrillo2019testing}. Other methods can include using data augmentation to boost the size of a dataset. In preparation for upcoming surveys and telescopes a great deal of work has been done to create automated lens finding tools to make the process of finding gravitational lenses in this large amount of data more effective. Within the \textit{Euclid} telescope consortium, there have been strong gravitational lens challenges, in which participants used various methods to classify simulated images as lenses and non-lenses \citep{metcalf2019strong}. Machine learning technologies deployed on lens finding include \correctionsbold{SVMs \citep{hartley2017support}}, autoencoders  \citep{cheng2020identifying}, transfer learning \citep{hezaveh2017fast}, ResNets \citep{lanusse2017cmu,huang2019finding, petrillo2019testing}, and CNNs \citep{pearson2018auto}. The most common architecture used are CNNs. These range from the very simplistic to the very complex.

Nevertheless, this technology still presents the community with a fundamental interpretability problem. How exactly are these lens-finding algorithms working? What features of the lenses are they learning and responding to? What categories of gravitational lens will they be good at finding, and are there strong lens categories or configurations that would be systematically missed by these machine learning algorithms? A naive and traditional approach would be to regard machine learning, and CNNs in particular, as black boxes whose performance can just be assessed with simulated inputs. Such an approach would only be as good as the extent to which simulations reproduce real data, and ``it just works'' is not usually considered a sufficient level of insight in the physical sciences. Moreover, awareness of inadvertent algorithmic biases is increasingly seen as a core problem in machine learning \citep[e.g.][]{obermeyer_ai_bias}, and tools are now being developed to shed light on the internal workings of these ostensible black box algorithms.

One does not need to dig very deep to discover that interpretability is a non-trivial problem. Deep learning models often rely on many millions of trainable parameters that depend on each other in very complicated ways. Historically, this complexity has made interpreting the operation of deep learning models very challenging. Moreover, CNNs can be sensitive to very subtle features of their input data, so images which look like noise to humans are able to generate high scoring results in CNNs \citep{nguyen2015deep}. Similarly, images of objects such as cats can be classified with high probability of being elephants by imposing an elephant skin texture over the image \citep{geirhos2018imagenet}. One of the major areas of research in machine learning at the current time is the development of techniques to interpret the results of machine learning algorithms. There are several ways in which machine learning is analysed to develop an understanding of how the algorithm is behaving, such as tactically changing the input values in a known way and evaluating how the output changes \citep{zeiler2014visualizing}. Other methods include using backpropagation to update the input image based on a target output value rather than the model weights. This updates an image to strongly activate the target class, which can reveal information that the model has learnt such as making forearms when trying to create an image of a dumbbell \citep{mordvintsev2015inceptionism}. 

There are various methods that attempt to highlight areas of the image that influence the decision making of the CNN. The simplest of these approaches is the generation of saliency maps \citep{simonyan2013deep}, including occlusion maps \citep[see section \ref{sec:occlusion} and][]{zeiler2014visualizing} and \textit{Class Activation Maps} \citep[see section \ref{sec:guided_grad_cam} and][]{selvaraju2017grad} like GradCAM and guided GradCAM. 
Alternative approaches to interpreting deep networks include deriving \textit{Class-Generated Images} (CGIs) using tools like Google's Deep Dream \citep{mordvintsev2015inceptionism}.

Understanding the decision making processes of machine learning algorithms is arguably the next important advance in the application of machine learning to astronomy. If one can understand the decisions the algorithms are making then one might be able to directly infer which image features or astrophysical phenomena  the algorithms may be selecting for or actively selecting against. This will help  quantify which categories of gravitational lenses that the algorithms can find and those they cannot. For example,  one of the possible categories of gravitational lens that could be missed by machine learning are source galaxies that are not blue, particularly if trained on existing surveys such as SLACS \citep{bolton2008sloan} which pre-selects early-type foreground galaxies and star-forming emission line background systems. The machine learning methods could, for example, be focusing on colour more than the general morphology of the lens system. \correctionsbold{CNNs have been shown to perform worse on galaxy-galaxy lenses that differ from the colour and PSF of lenses from their training data \citep{jacobs2022exploring}.} If one does not investigate which rare lens configurations the machine learning methods do not detect, there is a risk of losing interesting and important gravitational lens systems in the seas of data generated by future telescopes, as well as drawing spurious conclusions about the populations of sources and lenses, and drawing misleading cosmological inferences from them. Indeed there is already a precedent for human volunteers finding a strong lens with a red background source that would be missed by a SLACS-like automated lens finder \citep{geach2015}.

In this paper, we present new CNNs trained for finding strong gravitational lenses, and assess the networks' characteristics and performance with a wide variety of these interpretability tools. As a critical test case, we examine in particular whether the exotic subclass of compound lenses is detected by our CNNs, and attempt to determine why.  
In section 2, we describe the three different datasets we use within this paper. In section 3 we describe the network architecture and training process for the CNNs used within this work. Section 4 describes the results of the CNNs on the 3 different datasets. Section 5 investigates the interpretability of the CNNs and what they interpret as a lens and non-lens. Section 6 gives a discussion on this paper and section 7 is the conclusion.
A flat $\Lambda$CDM cosmology is used throughout with $H_0 = 70$\,km\,s$^{-1}$\, Mpc$^{-1}$ and matter density $\Omega_\mathrm{m}=0.3$.

\section{Data}\label{sec:data}
\subsection{Single Lenses}\label{sec:single_lenses}

    \begin{figure}
	\includegraphics[width=\columnwidth]{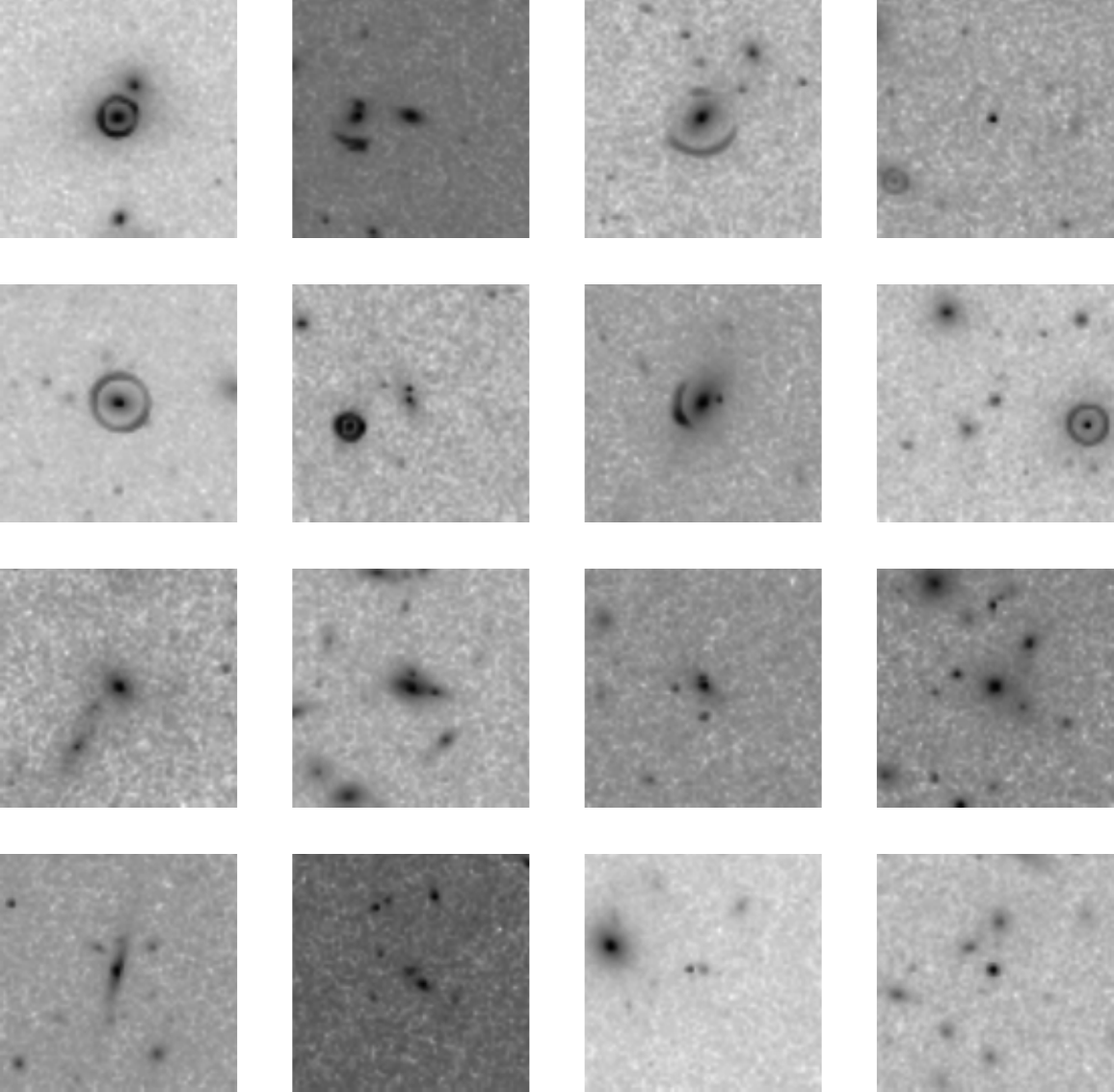}
    \caption{Log scale VIS band images from the training set. Top two rows show simulated single lenses. Bottom two rows show simulated non-lenses}
    \label{fig:SGLC2_Data}
\end{figure}

There is no empirical data set of the size of Euclid's and with its image characteristics, so training on simulations is inevitable. 
    The data used in this paper are simulated Euclid images from the VIS and NISP instruments, produced by the Bologna Lens Factory (B. Metcalf, priv. comm.). There are 100\,000 images in the training set and an additional 100\,000 images in the test set. The training set consists of \textasciitilde 72\,000 lenses and \textasciitilde 28\,000 non-lenses. The test set consists of \textasciitilde 71\,000 lenses and \textasciitilde 29\,000 non-lenses. The 
    NISP data consists of 3 bands: Y, J and H, while the VIS imaging is with a very broad band visual filter \citep{laureijs2011euclid}. These images are $66\times 66$ pixels with a pixel scale of 0.3 arcsec/pixel. The VIS images are $200 \times 200$ pixels with a pixel scale of 0.1 arcsec/pixel. \correctionsbold{An example of VIS images from this dataset are shown in figure \ref{fig:SGLC2_Data}. }

    The gravitational lens can be anywhere within the image. The point spread function (PSF) has an approximately three-fold symmetry  \citep{schmitz2020euclid}. The images are scaled linearly between 0 and 1 before being input to the CNNs.

\begin{figure}
	\includegraphics[width=\columnwidth]{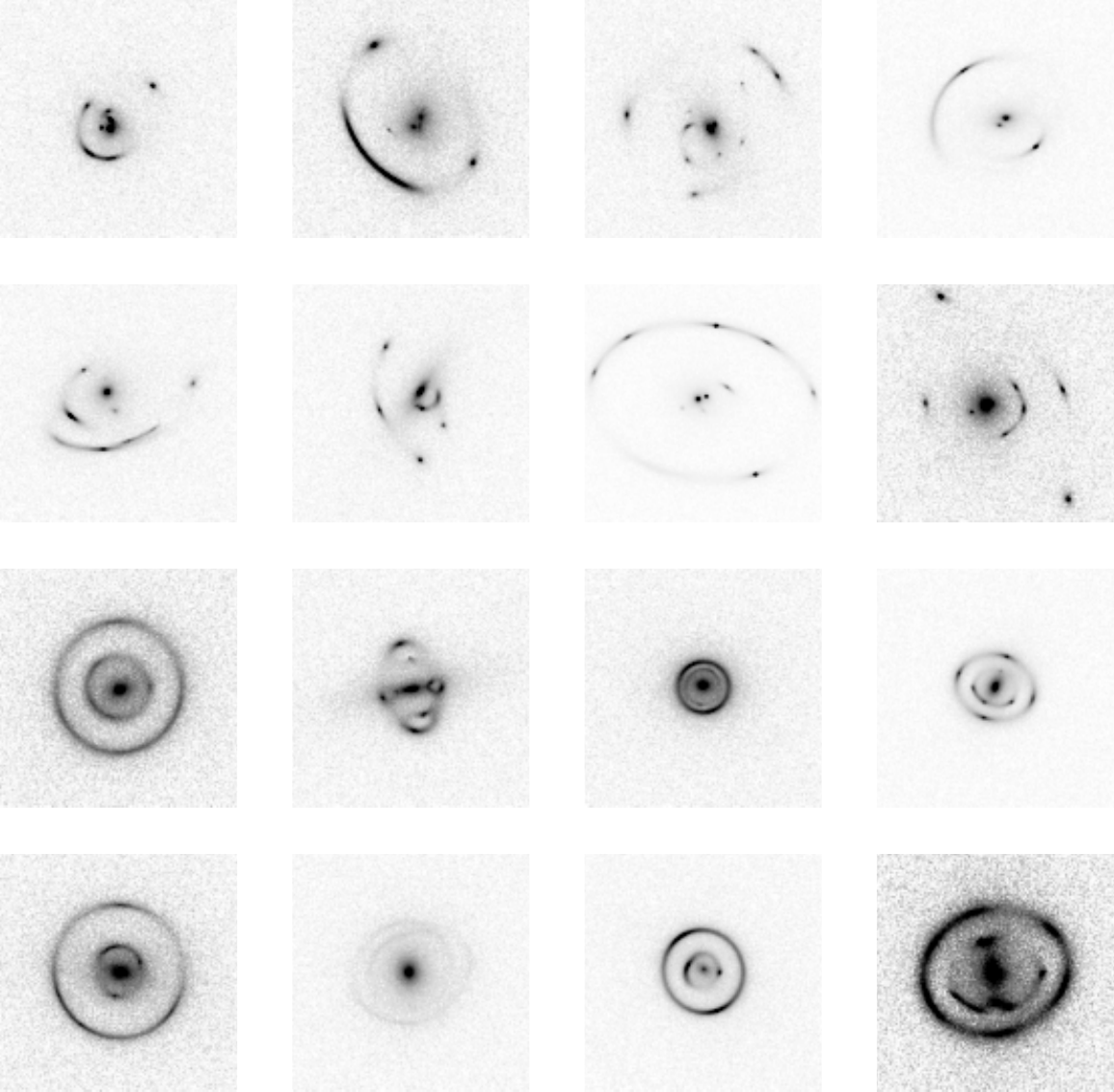}
    \caption{VIS band images of compound lenses. Top two rows show simulated compound arcs. Bottom two rows show simulated double rings}
    \label{fig:CompoundLens_square_Greys}
\end{figure}
    
\subsection{Compound Lenses}\label{sec:compound_lenses}
The simulated systems in section \ref{sec:single_lenses} did not contain compound lenses, and very few compound lensing systems have been discovered to date, so further simulations were required. 
Simulated images containing compound lenses were created using the \texttt{Lenstronomy} python package \citep{birrer2018lenstronomy} and \texttt{SkyPy} \citep{amara2021skypy}. Redshifts and magnitudes for the lens and source galaxies were randomly sampled in pairs from set of parameter values that were used to generate the single-lens training data. The source was sampled from source magnitude and source redshift distributions and the two lens galaxies were sampled from lens magnitude and lens redshift distributions. From these sampled magnitudes the galaxy velocity dispersions were calculated using the Faber-Jackson relationship \citep{faber1976velocity} and finally the Einstein radius of each lens were calculated using equation \ref{eq:vel2thetaE}: 
\begin{equation}
\theta  = 4\pi \left (\frac{\sigma_\mathrm{v}}{c}\right)^2 \frac{D_\mathrm{ls}} {D_\mathrm{s}}
\label{eq:vel2thetaE}
\end{equation}
where $\theta$ is the Einstein radius, $D_\mathrm{ls}$ is the lens--source angular diameter distance, $D_\mathrm{s}$ is the observer--source angular diameter distance, and $\sigma_\mathrm{v}$ is the lensing galaxy velocity dispersion. The other parameters of the galaxies were sampled from a distribution of SLACS lens data \citep{bolton2008sloan,auger2009sloan,newton2011sloan,shu2017sloan,denzel2021new}. A uniform distribution was used to assign these SLACS variables to the galaxies; these parameters are shown in table \ref{tab:Params_SLACS}. The lower redshift lens is more massive than the higher redshift lens two thirds of the time. The NISP bands have a simulated exposure time of 264s and the VIS band have a simulated exposure time of 1800s to match the single lens data for \textit{Euclid}-Wide. To approximate the \textit{Euclid} instruments' imaging characteristics, the simulated images were convolved with gaussian PSFs with full-widths at half maximum appropriate for the VIS and NISP instruments \citep{laureijs2011euclid}.

 \begin{table*}
	\centering
	\caption{Parameters used to generate simulated compound lenses.}
	\label{tab:Params_SLACS}
	\begin{tabular}{lccr} 
		\hline
		 Parameter & Lower Bound & Upper Bound & Units \\
		 \hline
		R Sersic & 0.2 & 7 & arcseconds\\
		n Sersic & 1.5 & 20 & - \\
		X Position of Source & -5 & 5 &  arcseconds\\
		Y Position of Source & -5 & 5 & arcseconds\\
		Angle & 0 & $\pi$ & radians\\
		Axis Ratio & 0.3 & 1 & -\\
		X Position of Compound Arc Lens galaxy & X Position of Source -5 & X Position of Source + 5 & arcseconds\\
		Y Position of Compound Arc Lens galaxy & Y Position of Source -5 & Y Position of Source + 5 & arcseconds\\
		X Position of Double Ring Lens galaxy & X Position of Source - 0.25 & X Position of Source + 0.25 & arcseconds\\
		Y Position of Double Ring Lens galaxy & Y Position of Source - 0.25 & Y Position of Source + 0.25 & arcseconds\\
		Gamma & -0.5 & 0.5 & - \\
		
		\hline
	\end{tabular}
\end{table*}

Two datasets of 10\,000 images were created. One used parameters that produce a set of compound lenses that are mainly arcs, while the other used a parameter set with tighter constraints on the galaxy positions generating mostly double Einstein rings (see table \ref{tab:Params_SLACS}). \correctionsbold{An example of these simulated compound lenses can be seen in figure \ref{fig:CompoundLens_square_Greys}.}

\subsection{Known Compound Lenses}\label{sec:euclidising}
\begin{figure}
	\includegraphics[width=\columnwidth]{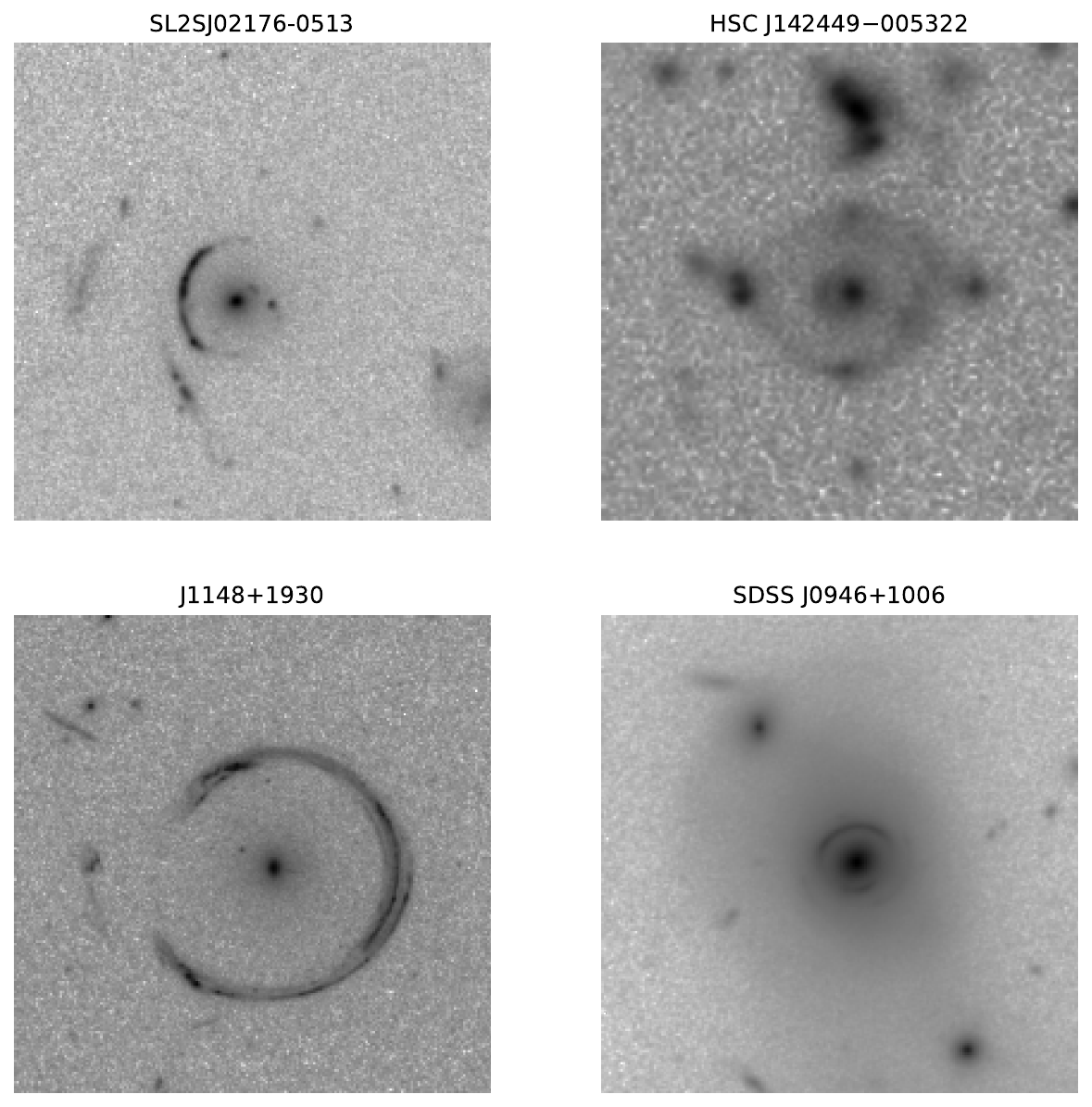}
    \caption{F814W band images of the four known compound lenses.}
    \label{fig:CompoundLens_square_BW}
\end{figure}

The CNNs described in this paper were applied to four known compound lenses SL2SJ02176-0513 \citep{2009A&A...501..475T}, J1148+1930 \citep{schuldt2019inner}, and SDSS J0946+1006 \citep{gavazzi2008sloan} from the Hubble Space Telescope (HST) and HSC J142449-5322 \citep{tanaka2016spectroscopically} from Hyper Suprime-Cam (HSC). The archival F814W-band images for these compound lenses are shown in figure \ref{fig:CompoundLens_square_BW}. A 20 $\times$ 20 arcsec postage stamp was extracted with the compound lens centred in the image. The pixel scales of these compound lens postage stamps were resampled to the same pixel scale as the training data, 0.1 arcsec/pixel for the bluest band and 0.3 arcsec/pixel for the rest. These data differ from the data on which the CNNs have been trained. These known compound lenses do not necessarily have 4 bands of imaging available. The CNN architectures require 4 bands of input data, so missing bands were populated by duplicating data from adjacent bands. The bands used for each compound lens are shown in table  \ref{tab:compoundLens_irl_Bands}. Each image was normalised between 0 and 1 prior to being input into the models. Each band was resampled to the pixel dimensions of the CNN as described in section \ref{sec:network_design}. 

 \begin{table}
	\centering
	\caption{Known compound lenses image bands to expected CNN bands.}
	\label{tab:compoundLens_irl_Bands}
	\begin{tabular}{lcccr} 
		\hline
		 Lens Name & VIS & Y & J & H \\
		 \hline
		SL20S J02176-0513 & F606 & F606 & F814 & F814 \\
		HSC J142449-005322 & R & I & Z & Y \\
		SDSS J1148+1930 & F475 & F606 & F606 & F814 \\
		SDSS J0946+1006 & F160 & F336 & F438 & F814\\
		\hline
	\end{tabular}
\end{table}

\section{Network Design}\label{sec:network_design}
\begin{figure}
	\includegraphics[width=\columnwidth]{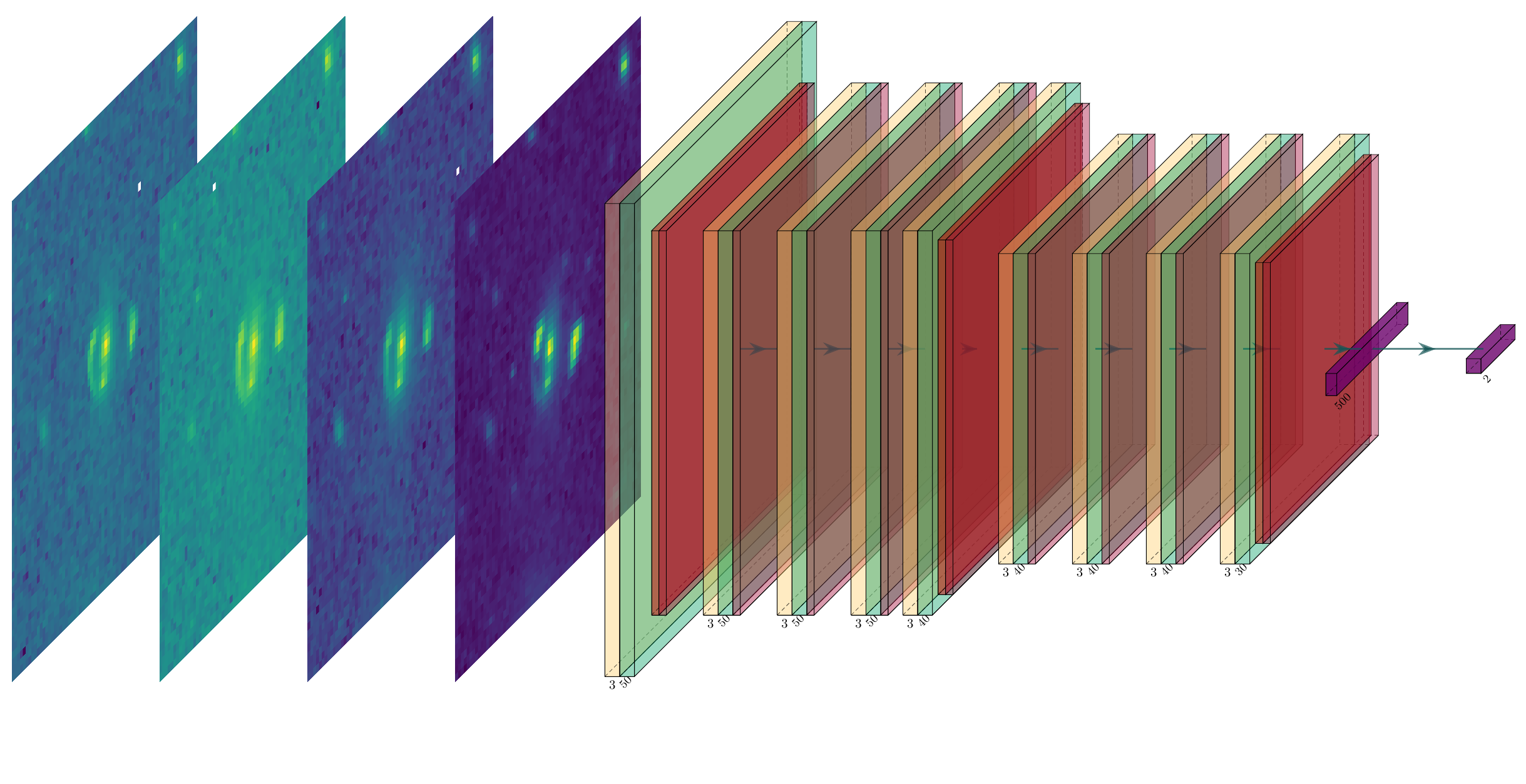}
    \caption{The architecture of OU-66. Yellow squares correspond to Conv2D layers, teal squares correspond to ReLU layers, orange layers correspond to MaxPool2D layers, and red squares correspond to Dropout layers.}
    \label{fig:OU-66}
\end{figure}

\begin{figure}
	\includegraphics[width=\columnwidth]{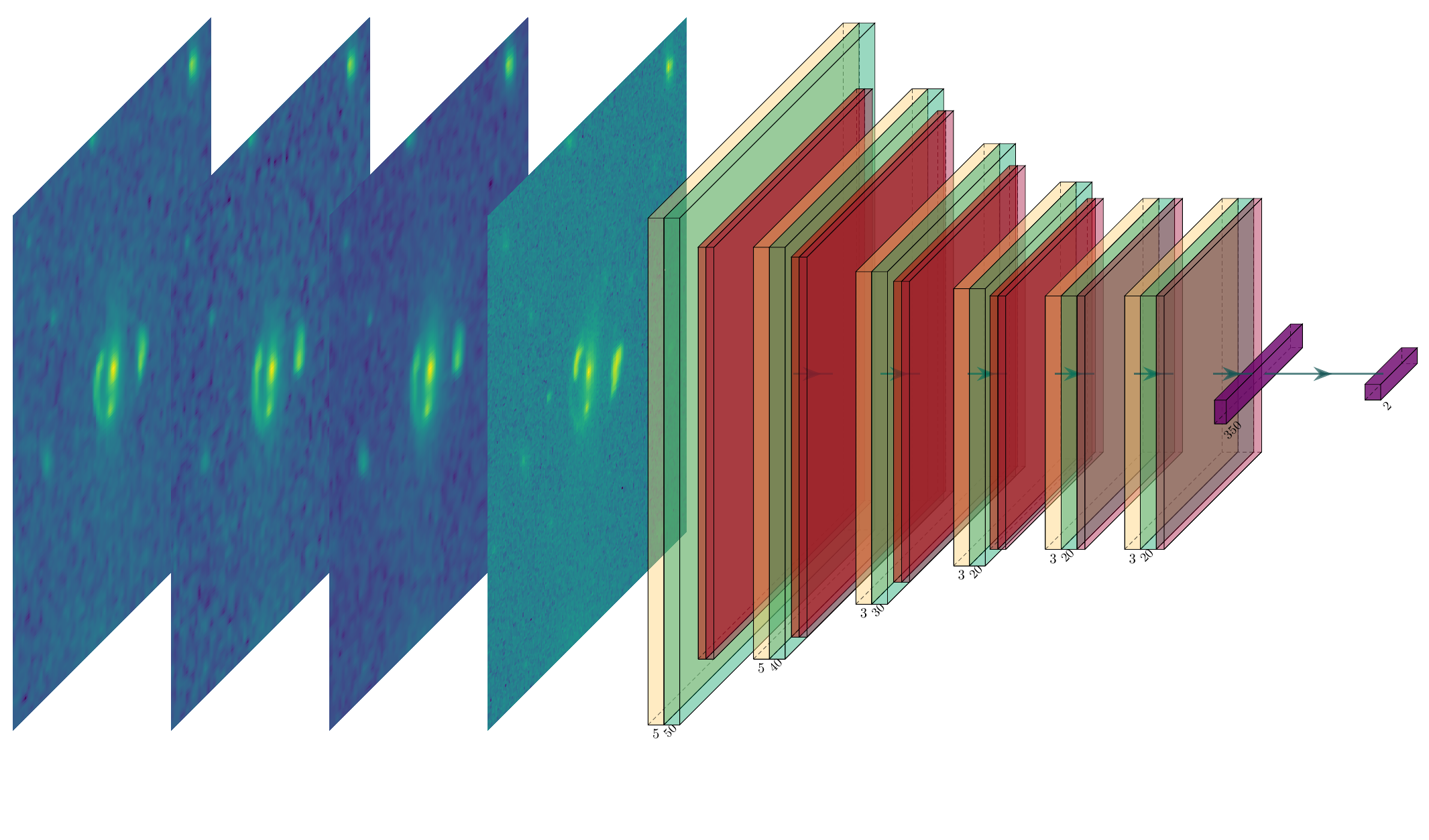}
    \caption{The architecture of OU-200. Yellow squares correspond to Conv2D layers, teal squares correspond to ReLU layers, orange layers correspond to MaxPool2D layers, and red squares correspond to Dropout layers.}
    \label{fig:OU-200}
\end{figure}

In this section details the creation, training, and testing of three CNN models; OU-66, OU-200, and OU-T-SNE (OU stands for Open University in this case). The architectures for both of these CNNs can be seen in tables \ref{tab:OU-66-Arc} and \ref{tab:OU-200-Arc} and graphical representations shown in figures \ref{fig:OU-66} and \ref{fig:OU-200}. 
\subsection{OU-66, OU-200 and variants}
These CNNs were created and trained using the Python package \texttt{PyTorch} \citep{pytorch}. Using these model designs a total of seven CNNs were created for this task, each accepting different configurations of input data. These CNNs can be segregated into two architectures depending on the pixel dimensions of the input data, if the data was 66 $\times$ 66 pixels the OU-66 architecture was used, if the data was $200 \times 200$ pixels the OU-200 architecture was used.  These architectures are outlined in tables \ref{tab:OU-66-Arc} and \ref{tab:OU-200-Arc}, and are based upon a CNN designed and trained on a similar problem by \citep{Davies_2019}.
Changes were made to this model including adding regularisation with dropout \citep{srivastava2014dropout}, adding another fully connected layer,   replacing $15 \times 15 $ convolutional kernels with two $5 \times 5$ convolutional kernels and replacing $5 \times 5$ convolutional kernels with two $3 \times 3$ convolutional kernels. The replacement of single large convolutional kernels with two smaller convolutional kernels is designed to reduce the computational time required, but the transformations being made are equivalent. \correctionsbold{The ReLU activation function \citep{agarap2018deep} is used after each convolutional layer this returns zero if the input is negative otherwise returns the input value.}
The way that inputs were handled was also changed by scaling them to be between 0 and 1, rather than normalising them to the maximum pixel value in each image.

Seven CNNs are OU-J, OU-Y, OU-H, OU-JYH, OU-VIS, OU-66, and OU-200 \correctionsbold{are used to create a single classifer OU-T-SNE in section \ref{sec:OU_T_SNE}}. These CNNs take input from the bands in their names; OU-66 and OU-200 take all 4 bands and resize the input to the corresponding size in their name, while the variations take only subsets of the bands. CNN variations that use near-infrared data only are based on OU-66, while OU-VIS is based on OU-200. The outputs of these models are used to generate OU-T-SNE. 

\subsection{CNN Training}
Each CNN is initialised with He Normalisation \citep{he2015delving}. Images from each band were combined into a single input tensor with dimensions shown in tables \ref{tab:OU-66-Arc} and \ref{tab:OU-200-Arc}. The CNNs were trained for 250 epochs with a learning rate\footnote{The learning rate is a dimensionless scale factor that determines the step size in the optimisation.} of $3 \times 10^{-4}$ using the Adam optimiser \citep{kingma2014adam} and a batch size of 250 for OU-66, and 125 for OU-200. They were each trained on the Open University GPU cluster\footnote{Node specification: two INTEL XEON GOLD 5118 processors with 12 cores of 24 threads (2.30 GHz); 3D controllers NVIDIA Corporation GP100GL (Tesla P100 PCIe 12GB); CPU memory 251 GB.} for approximately seven hours.

The Euclid Strong Lensing data challenge used two distinct datasets. The first of these sets (COMP) contains 60\,000 images and was provided to participants to allow them to design, train, refine and validate their challenge entries. The second dataset (EVAL) contains 100\,000 images that were reserved for final evaluation of the individual challenge entries. While designing the CNNs in this paper, the COMP dataset was further subdivided into three subsets -- the ``training set'', the ```validation set'' and the ``COMP test set''. The training set contains 45\,000 images, the validation set contains 3\,000 images and the COMP Test set contains 12\,000 images. If the CNN took input from both the VIS and the NISP bands, the bands were resampled to the input size of the CNN using the resize function from the python package \texttt{scikit-image} \citep{scikit-image}. Attempting to balance the non-lens class by rotating images caused the CNN to achieve spurious accuracies greater than 90\% by learning the orientation of the PSF, despite the asymmetry of the PSF being not discernable by eye in the images. 
To avoid this problem the class imbalance was mitigated by applying a weighting to the loss function. \correctionsbold{Categorical cross entropy loss is shown in equation \ref{eq:catloss} where M is the number of classes, y is 1 if the class label (c) is the correct classification for the image (i), p is the predicted value for the image (i) in class (c).} A categorical cross entropy loss was used instead of a binary classifier to allow the CNN to generate schemata for both lenses and non-lens. In contrast, training a binary classifier will only yield an internal representation or schema for its target class i.e. \textit{either} for lens \textit{or} non-lens. The training data was shuffled during training.

\begin{equation}
\mathrm{loss} = -\sum_{c=1}^M y_{i,c} \ln(p_{i,c})
\label{eq:catloss}
\end{equation}

\begin{figure}
	\includegraphics[width=\columnwidth]{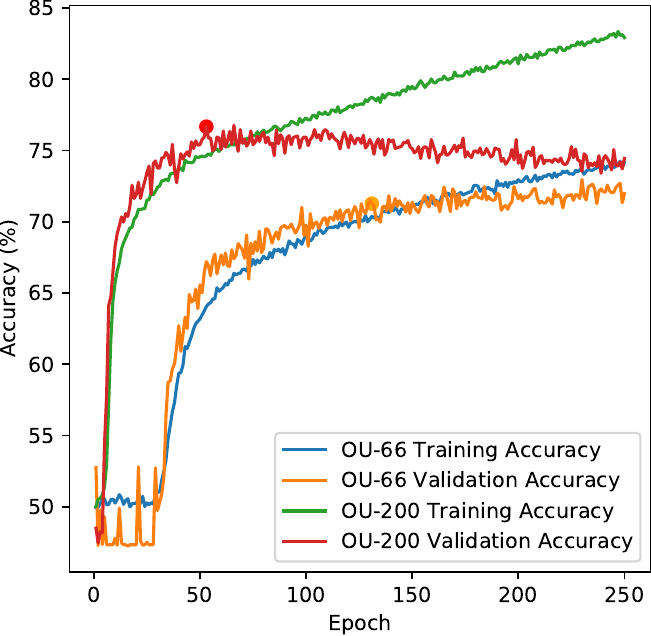}
    \caption{The Accuracy of the training and validation data over the course of training. The final recorded weights are shown with a dot}
    \label{fig:TrainingCurveAcc}
\end{figure}

\begin{figure}
	\includegraphics[width=\columnwidth]{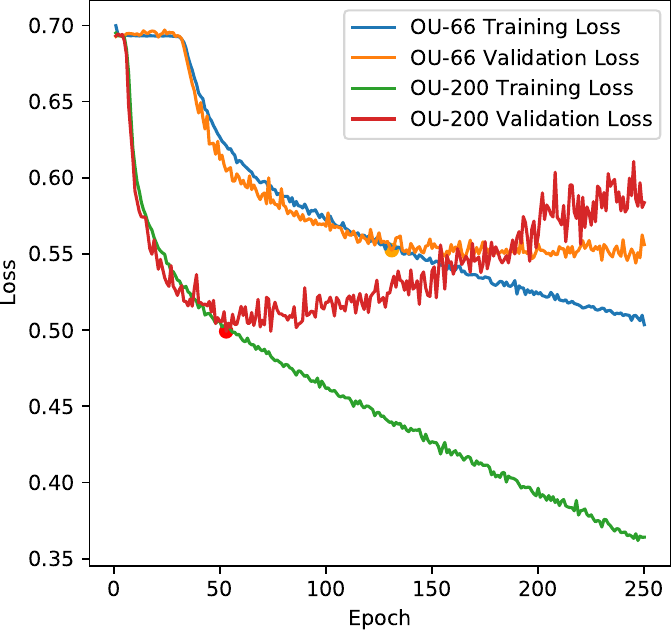}
    \caption{The Loss of the training and validation data over the course of training. The final recorded weights are shown with a dot}
    \label{fig:TrainingCurveLoss}
\end{figure}

\begin{figure}
	\includegraphics[width=\columnwidth]{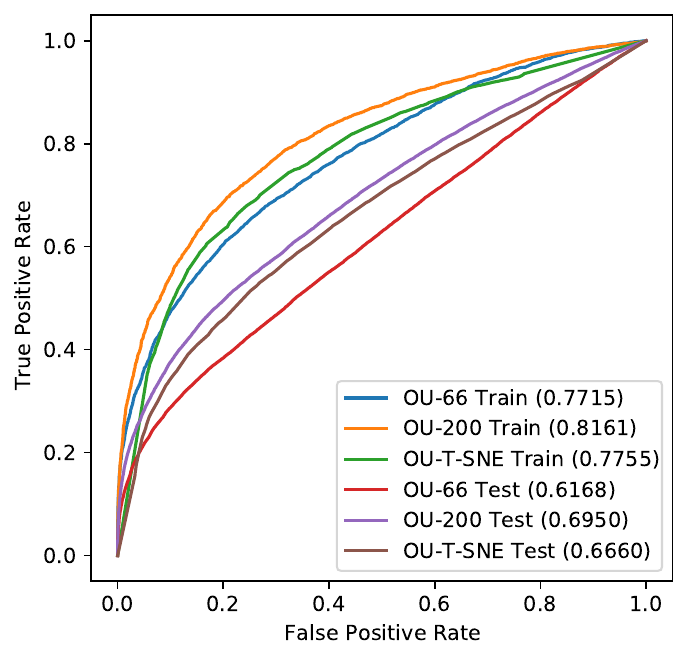}
    \caption{ROC Curve for each model on the training and testing data.}
    \label{fig:ROC}
\end{figure}

During training the time taken for each epoch, the epoch number, the training loss, the training accuracy, the validation loss, and the validation accuracy were recorded. \correctionsbold{The parameters show the learning process of the CNN. Eventually the CNNs are expected to have memorised the entire training set. This leads to the training accuracy increasing over time until it memorises the entire dataset. At this stage in CNN can often no longer generalise to unseen data. The validation dataset is not seen whilst the CNN is training this acts as a proxy for the CNN's ability to generalise to unseen data. The training and validation loss represent the same concepts in terms of categorical cross entropy loss instead of accuracy. The CNN weights for each trained model are for the epoch which has the lowest loss.} The evolution of the training and validation accuracies during training are shown in figure \ref{fig:TrainingCurveAcc}, and the corresponding loss evolution is shown in figure \ref{fig:TrainingCurveLoss}. The model weights were saved whenever the model reached a new minimum validation loss. 
The Receiver Operating Characteristic (ROC) curves for these models are shown in figure \ref{fig:ROC}.

 \begin{table*}
	\centering
	\caption{OU-66 architecture. The ReLU activation functions are considered layers in this table and follow all Conv2D layers.}
	\label{tab:OU-66-Arc}
	\begin{tabular}{lccccccccccr} 
		\hline
		Layer & Type & Kernel/Pool & Kernel/Neuron & Stride & Padding & Activation/ & Input shape &  Output Shape & Parameter \\
		& & Size & Count & & & Dropout Probability & & & Count\\
		\hline
		0 & Conv2D & (3,3) & 50 & (1,1) & Valid & ReLU & (4,66,66)  & (50,64,64) & 1\,850\\
		2 & MaxPool2D & (2,2)  & - & (1,1) & - & - & (50,64,64) & (50,32,32) & 0\\
		3 & Dropout & - & - & - & - & 0.2 & (50,32,32)  & (50,32,32) & 0\\
		4 & Conv2D & (3,3) & 50 & (1,1) & Valid & ReLU & (50,32,32)  & (50,30,30) & 22\,550\\
		6 & Dropout & - & - & - & - & 0.2 & (50,30,30) & (50,30,30) & 0\\
		7 & Conv2D & (3,3) & 50 & (1,1) & Valid & ReLU &  (50,30,30) & (50,28,28) & 22\,550\\
		9 & Dropout & - & - & - & - & 0.2 & (50,28,28)  & (50,28,28) & 0\\
		10 & Conv2D & (3,3) & 50 & (1,1) & Valid & ReLU & (50,28,28)  & (50,26,26) & 22\,550\\
		12 & Dropout & - & - & - & - & 0.2 & (50,26,26) & (50,26,26) & 0\\
		13 & Conv2D & (3,3) & 40 & (1,1) & Valid & ReLU & (50,26,26)  & (40,24,24) & 18\,040\\
		15 & MaxPool2D & (2,2)  &- & (1,1) & - & - & (40,24,24)  & (40,12,12) & 0\\
		16 & Dropout & - & - & - & - & 0.2 & (40,12,12)  & (40,12,12) & 0\\
		17 & Conv2D & (3,3) & 40 & (1,1) & Valid & ReLU & (40,12,12)  & (40,10,10) & 14\,440\\
		19 & Dropout & - & - & - & - & 0.2 & (40,10,10)  & (40,10,10) & 0\\
		20 & Conv2D & (3,3) & 40 & (1,1) & Valid & ReLU & (40,10,10)  & (40,8,8) & 14\,440\\
		22 & Dropout & - & - & - & - & 0.2 & (40,8,8)  & (40,8,8) & 0\\
		23 & Conv2D & (3,3) & 40 & (1,1) & Valid & ReLU & (40,8,8) & (40,6,6) & 14\,440\\
		25 & Dropout & - & - & - & - & 0.2 & (40,6,6)  & (40,6,6) & 0\\
		26 & Conv2D & (3,3) & 30 & (1,1) & Valid & ReLU & (40,6,6)  & (30,4,4) & 10\,830\\
		28 & MaxPool2D &  (2,2)  & - & (1,1) & - & - & (30,4,4)  &(30,2,2) & 0\\
		29 & Dropout & - & - & - & - & 0.2 & (30,2,2) & (30,2,2) & 0\\
		30 & Dense & - & 500 & - & - & ReLU & 120  & 500 & 60\,500\\
		32 & Dense & - & 2 & - & - & Softmax & Input shape & 2 & 1\,002\\
		\hline
		\multicolumn{2}{l}{Total Parameter Count} &  &  &  &  & &  &  & 203\,192\\
		\hline
	\end{tabular}
\end{table*}

\begin{table*}
	\centering
	\caption{OU-200 architecture. The ReLU activation functions are considered layers in this table and follow all Conv2D layers.}
	\label{tab:OU-200-Arc}
	\begin{tabular}{lccccccccccr} 
		\hline
		Layer & Type & Kernel/Pool & Kernel/Neuron & Stride & Padding & Activation/ & Input shape &  Output Shape & Parameter \\
		& & Size & Count & & & Dropout Probability & & & Count\\
		\hline
		0 & Conv2D & (5,5) & 50 & (1,1) & Valid & ReLU & (4,200,200) &(50,196,196)& 5\,050\\
		2 & MaxPool2D & (2,2) & -  & (1,1) & - & - & (50,196,196 & (50,98,98) & 0\\
		3 & Dropout & - & - & - & - & 0.2 & (50,98,98)& (50,98,98)& 0\\
		4 & Conv2D & (5,5) & 50 & (1,1) & Valid & ReLU & (50,98,98) & (40,94,94) & 50\,040\\
		6 & MaxPool2D & (2,2)  & - & (1,1) & - & - & (40,94,94) & (40,47,47)& 0\\
		7 & Dropout & - & - & - & - & 0.2 & (40,47,47)& (40,47,47)& 0\\
		8 & Conv2D & (3,3)& 40 & (1,1) & Valid & ReLU & (40,47,47 & (30,45,45) & 10\,830\\
		10 & MaxPool2D & (2,2) & - & (1,1) & - & - & (30,45,45) & (30,22,22)& 0\\
		11 & Dropout & - & - & - & - & 0.2 & (30,22,22) & (30,22,22)& 0\\
		12 & Conv2D & (3,3)& 30 & (1,1) & Valid & ReLU & (30,22,22) & (20,20,20) & 5\,420\\
		14 & MaxPool2D & (2,2) & - & (1,1) & - & - & (20,20,20) & (20,10,10)& 0\\
		15 & Dropout & - & - & - & - & 0.2 & (20,10,10) & (20,10,10)& 0\\
		16 & Conv2D & (3,3) & 20 & (1,1) & Valid & ReLU & (20,10,10) & (20,8,8) & 3\,620\\
		18 & Dropout & - & - & - & - & 0.2 & (20,8,8) & (20,8,8) & 0\\
		19 & Conv2D & (3,3) & 20 & (1,1) & Valid & ReLU & (20,8,8) & (20,6,6) & 3\,620\\
		21 & Dropout & - & - & - & - & 0.2 & (20,6,6) & (20,6,6)& 0\\
		22 & Dense & -& 350 & - & - & ReLU & 720  & 350 & 252\,350\\
		24 & Dense & - & 2 & - & - & Softmax & 350 & 2 & 702\\
		\hline
		\multicolumn{2}{l}{Total Parameter Count} &  &  &  &  & &  &  & 331\,632\\

		\hline
	\end{tabular}
\end{table*}

\subsection{OU-T-SNE}\label{sec:OU_T_SNE}

    \begin{figure}
	\includegraphics[width=\columnwidth]{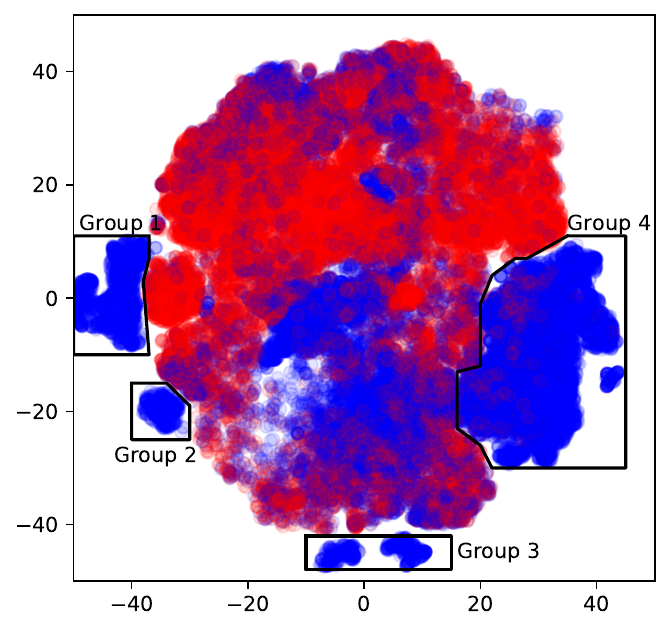}
    \caption{T-SNE embedding of the 7 different model outputs for the training set. This is a dimensionality reduction of the 7D array into 2D. The position of the image in this 2D reduction is used to classify the image as a lens or non-lens. Blue dots represent lenses and red represent non-lenses. }
    \label{fig:T-SNE_Embedding}
\end{figure}

\begin{figure}
	\includegraphics[width=\columnwidth]{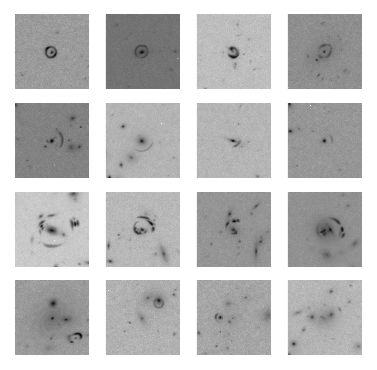}
    \caption{Examples of images within the groups shown in figure \ref{fig:T-SNE_Embedding}. Row 1: Images from Group 1, Row 2: Images from Group 2, Row 3: Images from Group 3, Row 4: Images from Group 4. }
    \label{fig:T-SNE_Examples}
\end{figure}

The OU-T-SNE model is a collection of 7 different models. These models have the same architectures as OU-66 \& OU-200 but with different inputs. There are 5 variations of OU-66, including  three where the input is a single NISP band, one with all the NISP bands, and finally the original OU-66 configuration. In addition, OU-200 and a variation of OU-200 where the input is a single-band VIS image are used. These 7 models are trained independently and the output of each model for the lens class is combined into a seven-element vector. 

The 7-dimensional combined model outputs were dimensionally reduced to 2D using a \textit{t}-distributed stochastic neighbour embedding (T-SNE) \citep{maaten2008visualizing}. This embedding is shown in figure \ref{fig:T-SNE_Embedding} where blue dots represent inputs containing simulated lenses and red dots represent non-lenses. Inspecting the members of the clusters within this embedding revealed that they represent different types of lenses and non-lenses. To classify the images within this embedding this space was divided into a $100 \times 100$ grid. Each square within the grid was assigned a classification value between zero and one based upon the fraction of lenses within the square. The test data were classified by deriving their locations in the embedded space and using the classification value assigned to the grid element at that location. In some cases, the embeddings for elements of the test set fell in grid elements that contained zero entries from the training data, and therefore no corresponding classification value. In such cases, the classification value was estimated using the mean of the eight surrounding grid values if they are defined. All of the test data within a square were assigned the same value for their classification strengths.

The T-SNE algorithm has clustered model predictions into a 2D space as shown in figure \ref{fig:T-SNE_Embedding} here four groups are highlighted. Examples of images in these groups are shown in figure \ref{fig:T-SNE_Examples}. Group one consists of mainly small centred Einstein rings. Group two mainly consists of arcs placed above, below, and to the right of the central galaxy. Group three mainly consists of gravitational lenses with multiple components. Group four mainly consists of off centred gravitational lenses.

\section{Results}
The real-valued model outputs were converted into binary classifications by thresholding. Any output value below 0.5 is classified as a non-lens and output values that are greater than or equal to 0.5 are classified as lenses.

Based on comparisons between the classification results and the known ground truths for each of the EVAL set images, the performances of the models were evaluated by maximising the $F_\beta$ score. The $F_\beta$ score \citep{chinchor1992muc}, defined as
\begin{equation}
F_\beta = \frac{(1+\beta^2) \times (\mathrm{precision} \times \mathrm{recall})}{(\beta^2 \times \mathrm{precision} + \mathrm{recall})}
\label{eq:Fbeta}
\end{equation}
is a weighted average of precision, defined as 
\begin{equation}
\mathrm{Precision} = \frac{\mathrm{True\: Positives}}{\mathrm{True\: Positives} + \mathrm{False\: Positives}}
\label{eq:precison}
\end{equation}
recall defined as, 
\begin{equation}
\mathrm{Recall} = \frac{\mathrm{True\: Positives}}{\mathrm{True\: Positives} + \mathrm{False\: Negatives}}
\label{eq:recall}
\end{equation}
and accuracy defined as
\begin{multline}
\mathrm{Accuracy} = \\ \frac{\mathrm{True\: Positives} + \mathrm{True\: Negatives}}{\mathrm{True\: Positives} + \mathrm{True\: Negatives} + \mathrm{False\: Positives} + \mathrm{False\: Negatives}}{\mathrm ~.}
\label{eq:acc}
\end{multline}

The ROC curve for the training set and the EVAL set are shown in figure \ref{fig:ROC}. When the Area Under the ROC curve (AUROC) is 1 the model is a perfect classifier, when the AUROC is 0.5 the model is a random classifier. 

The goal for this challenge was to try to create a fairly complete (albeit slightly contaminated) lens sample, which could be the basis of follow-ups. 
Therefore, we used $\beta=0.03$ in model evaluation, which places greater weight on recall than precision. Other values of $\beta$ could be well-justified for other science goals, such as where the presence of false positives would represent a critical failure (e.g. JWST observations).

 \begin{table}
	\centering
	\caption{Metrics used to evaluate each model used in this paper, when the threshold is set to 0.5.}
	\label{tab:metric_table}
	\begin{tabular}{lccccr} 
		\hline
		 & $F_{\beta}$ & Precision & Recall & AUROC & Accuracy \\
		\hline
		OU-66 & 0.9727 & 0.9758 & 0.2180 & 0.617 & 0.3308\\
        OU-200 & 0.9687 & 0.9697 & 0.4423 & 0.695 & 0.5143\\
		OU-T-SNE & 0.9423 & 0.9431 & 0.4846 & 0.666 & 0.5371\\
		\hline
	\end{tabular}
\end{table}

The datasets described in Section \ref{sec:data} were classified by the models OU-66, OU-200, and OU-T-SNE. The metrics used to evaluate these models on the EVAL dataset are shown in \correctionsbold{table} \ref{tab:metric_table} and the recalls of these models on all the datasets used in this paper are shown in table \ref{tab:compoundLens_Results}. 
The differences in recall between the single lenses and the compound lenses could be due to the differences between the training data and the compound lens data. The arcs and rings in the compound lens data tend to be redder than the training data, there are no red rings or arcs in the training data. The colour difference could be causing the model to struggle with some lens configurations. Moreover, the compound lens images do not contain any additional background galaxies unlike the those in the COMP and EVAL datasets.

 \begin{table*}
	\centering
	\caption{Recall of the models when the threshold is set to 0.5.}
	\label{tab:compoundLens_Results}
	\begin{tabular}{lccccccr} 
		\hline
		 & OU-H & OU-J & OU-Y & OU-JYH & OU-VIS & OU-66 & OU-200 \\
		 \hline
		Train Data & 0.4476 & 0.3636 & 0.2585 & 0.3170 & 0.2597 & 0.3807 & 0.4004  \\
		Test Data & 0.4772 & 0.3989 & 0.2752 & 0.3555 & 0.2495 & 0.2180 & 0.44.23 \\
		Arcs & 0.0706 & 0.5111 & 0.4083 & 0.7771 & 0.1684 & 0.6291 & 0.7505 \\
		Double Rings & 0.0908 & 0.5393 & 0.4667 & 0.7461 & 0.1618 & 0.4742 & 0.5567\\
		\hline
	\end{tabular}
\end{table*}

The model outputs for each of the known compound lenses are shown in table \ref{tab:compoundLens_irl_Results}.

 \begin{table}
	\centering
	\caption{Model predictions on known HST and HSC data.}
	\label{tab:compoundLens_irl_Results}
	\begin{tabular}{lcr} 
		\hline
		 Lens Name & OU-66 & OU-200 \\
		 \hline
		SL20S J02176-0513 & 0.9993 & 0.9852\\ 
		HSC J142449-005322 & 0.9302 & 0.7104 \\
		J1148+1930 & 1.0000 & 0.9953 \\
		SDSS J0946+1006 & 0.6873 & 0.8228\\
		\hline
	\end{tabular}
\end{table}

\section{Interpretability of the classifications}\label{sec:interpretability}
This section describes a set of investigations which were conducted in order to help explain how OU-66 and OU-200 make their decisions when classifying images. Five main techniques are used in this paper to develop an understanding of the model: 
\begin{enumerate}
    \item Covering parts of the image and recording how the model responds \citep[i.e. occlusion mapping, ][]{zeiler2014visualizing}.
    \item Visually inspecting the images which highly activate kernels within the model \citep{erhan2009visualizing}.
    \item Generating and inspecting images to highly activate either of the target classes \citep{yosinski2015understanding, simonyan2013deep}.
    \item Applying \textit{Deep Dream} \citep{mordvintsev2015inceptionism} to amplify features within simulated images that correspond to ``lens'' and ``non-lens'' features.
    \item Applying \textit{Grad-CAM} \citep{selvaraju2017grad} to highlight regions of input images that strongly activate for the ``lens'' class.
\end{enumerate}
In this section, the NISP bands are combined into an RGB image where red is the H-band, green is J-band, and blue is Y-band. Yellow is a summation of the H and J bands, cyan is a combination of J and Y bands, and magenta is a combination of H and Y bands. The majority of the code used to generate the representations shown in this section is adapted from \citep{uozbulak_pytorch_vis_2021}.

\subsection{Occlusion Maps}\label{sec:occlusion}
Occlusion mapping involves recording the change in models' responses to input images as some of their pixels are masked out. If the model output for the ``lens'' class using the masked image decreases relative to that when using its unmasked counterpart then it is more likely that the occluded pixels are associated with a lensing feature \citep{zeiler2014visualizing}.

In the case of OU-200 the mask was $4 \times 4$ pixels for OU-200, while for OU-66 it was $1 \times 1$ pixels. Model outputs were recorded with the mask placed at positions on a grid with 4 pixel intervals in the horizontal and vertical directions for OU-200 and 1 pixel intervals for OU-66\footnote{Different grid sizes were used for the two models because the inputs for OU-200 include resampled images from the NISP camera which have a native pixel size 3 times larger than that of VIS. Occluding fewer than 4 pixels in the OU-200 inputs would leave redundant information from the resampled NISP images unmasked.}.

The blue pixels in figures \ref{fig:Occ_SGLC2}, \ref{fig:Occ_Simulated}, and \ref{fig:CompoundLens_square_Occ} indicate areas where occlusion reduces the model output for the ``lens'' class, suggesting that the occluded pixels form part of a feature that is associated with lensing. Conversely, red pixels in figures \ref{fig:Occ_SGLC2}, \ref{fig:Occ_Simulated}, and \ref{fig:CompoundLens_square_Occ} indicate areas where occlusion increases the model output for the lens, suggesting the occluded pixels form part of a feature associated with the ``non-lens'' class. 

\subsubsection*{Single Lenses}
\begin{figure}
	\includegraphics[width=\columnwidth]{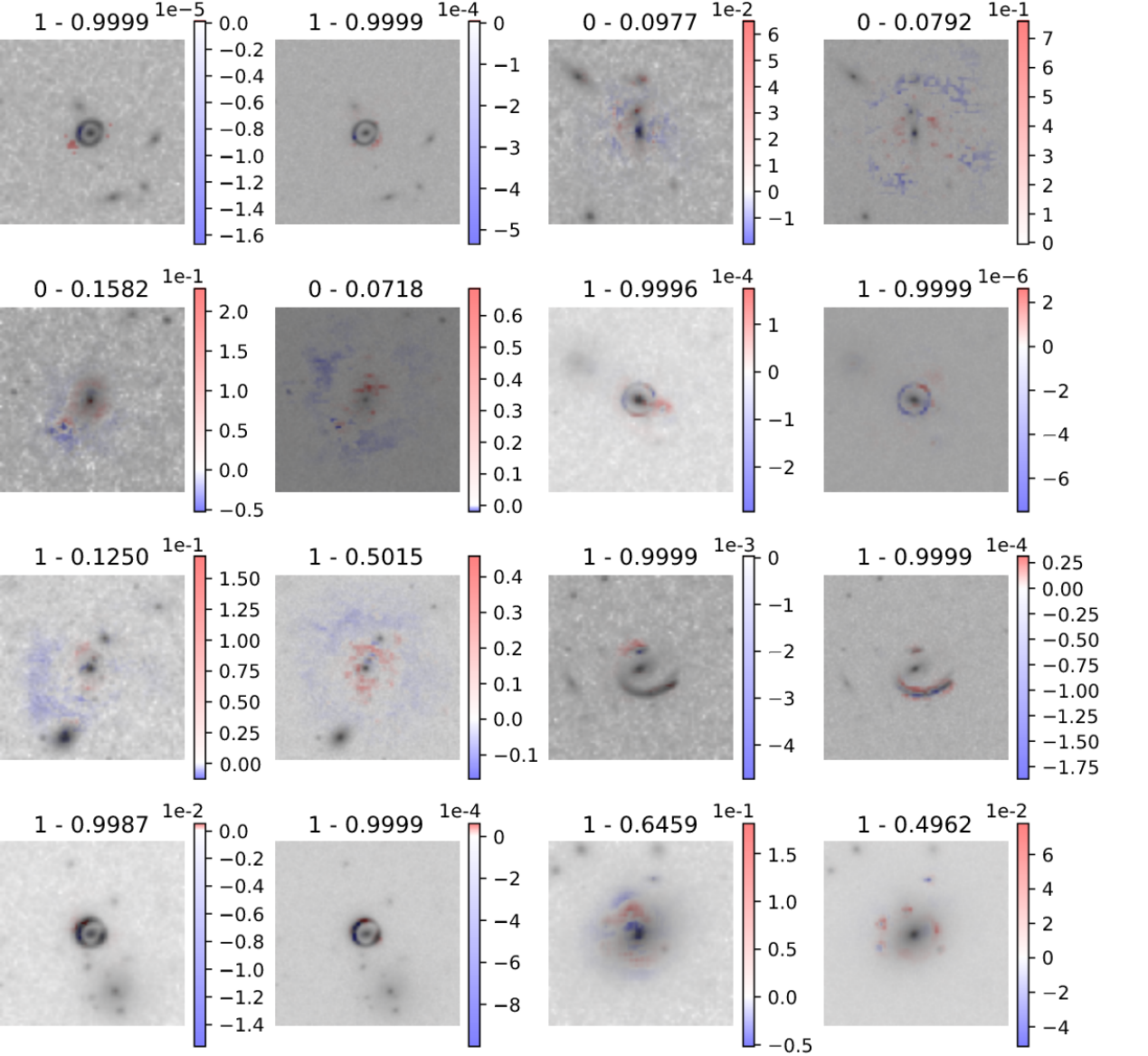}
    \caption{Occlusion Maps of SGLC2 data on OU-66 and OU-200. \correctionsbold{The first and third columns show the occlusion maps for OU-66, the second and fourth columns show the occlusion maps for OU-200 for the same images to their left. The left number above the image indicates if the image is a lens (1) or a non-lens (0) and the right number is the CNN output for that image.} The occlusion maps are independently scaled to highlight the change in output for each image, where the minimum value is blue, the maximum value is red, and the value of 0 is white. Blue indicates that the model associates the occulted feature with lensing and red indicates that the model associates the occulted feature with the non-lensing class. }
    \label{fig:Occ_SGLC2}
\end{figure}

Figure \ref{fig:Occ_SGLC2} shows the computed occlusion maps for both models (rescaled as necessary) overlaid on log-scaled VIS images. The top two rows show the maps for OU-66 and the bottom two rows show the maps for OU-200. 

For OU-66 and OU-200, the occlusion maps for images for which the models strongly predict the ``lens'' class tend to highlight the arcs and rings in the image in blue. This suggests that the model has learnt to associate rings and arcs with gravitational lensing. For images for which the models strongly predict the ``non-lens'' class, the main patterns seen are a red region in the centre of the image, surrounded by a faint disjointed blue ring roughly 10 arcseconds in diameter. It is possible that these fain blue rings arise because the training data contain a large number of ``lens'' images that have central lenses with roughly the same Einstein radius as these blue rings. This could result in the model trying to find a lens at this position within the noise of the data when a lens is not detected by the model. Spiral galaxy examples are also shown; the occlusion maps show arcs within the spiral arms, but these are identified as non-lens features by the model. This might mean that the model does not blindly associate arcs with lensing but also considers the context in which the arcs appear.

\subsubsection*{Compound Lenses}
\begin{figure}
	\includegraphics[width=\columnwidth]{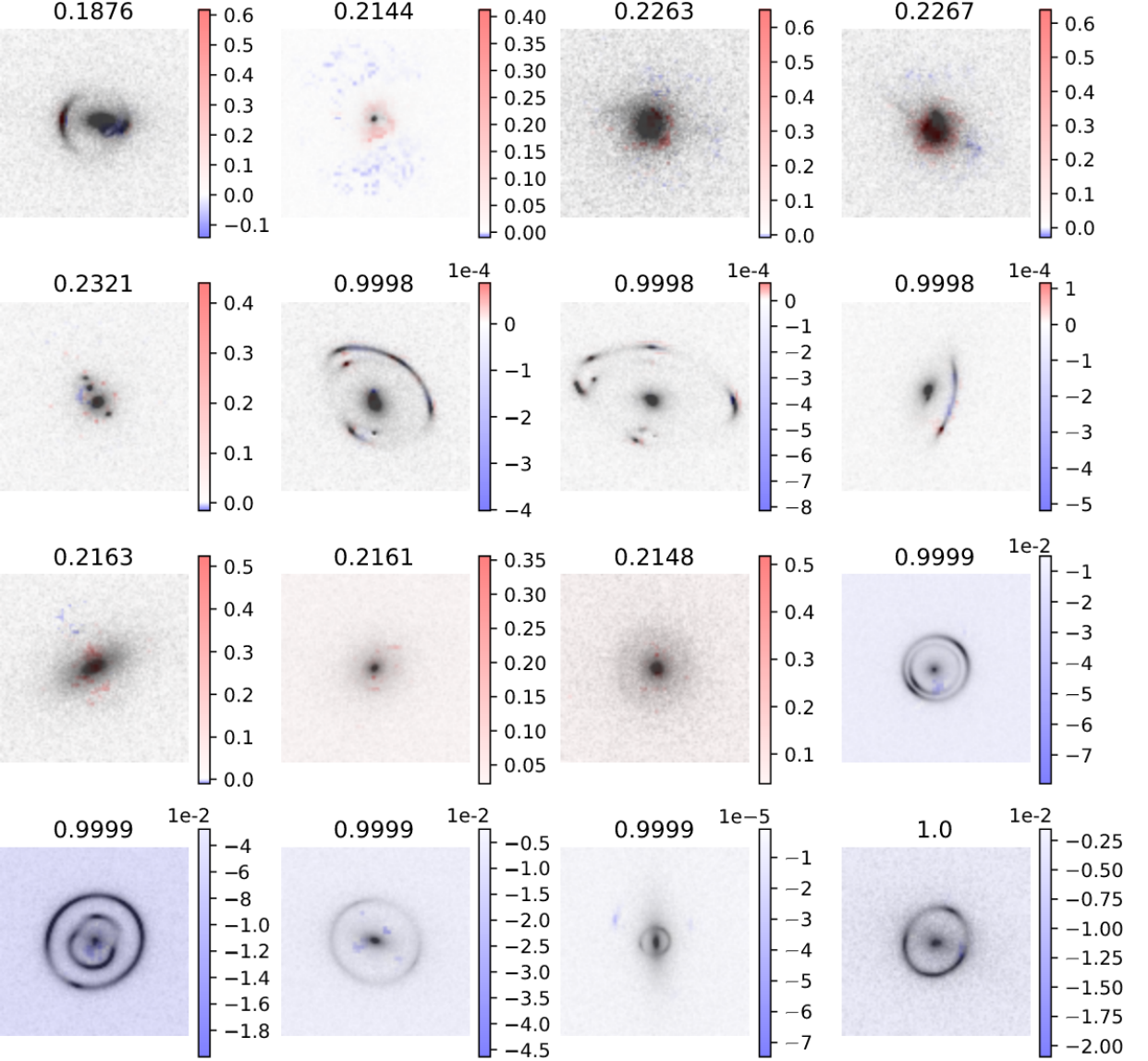}
    \caption{Occlusion Maps of simulated compound arcs and simulated double rings. The top 2 rows show the occlusion maps for simulated compound arcs and the bottom 2 rows show the occlusion maps for simulated double rings. The number \correctionsbold{above the image} is the model output for that image. The occlusion maps are independently scaled to highlight the change in output for each image, where the minimum value is blue, the maximum value is red, and the value of 0 is white. Blue indicates that the model associates the occulted feature with lensing and red indicates that the model associates the occulted feature with the non-lensing class. }
    \label{fig:Occ_Simulated}
\end{figure}

The occlusion maps for the simulated compound arcs are shown in the top two rows and double rings are shown in the bottom two rows of figure \ref{fig:Occ_Simulated}. 
The compound lens images for which the model predicts $\lesssim0.3$ for the ``lens'' class tend to have a central red ring or filled circle indicating that the model considers the pixels in this area to disfavour a lensing feature. Surrounding this area is a large blue region indicating that the model could be looking for rings and arcs at this diameter, which is similar to that of the faint blue rings shown in the ``non-lens'' images figure \ref{fig:Occ_SGLC2}. The images in this data for which the model predicts $\gtrsim 0.99$ are similar to the examples shown in \ref{fig:Occ_SGLC2}. The occlusion maps highlight lensing areas along the arcs and rings, with a strong non-lensing feature on the boundary of these arcs.

The double ring images for which the model predicts $\lesssim0.3$ for the ``lens'' class in figure \ref{fig:Occ_Simulated} tend to have strong non-lensing features in the centre of the image and there are occasionally low-level lensing features at the same radius as seen in the compound arc images. The images in this data for which the model predicts $\gtrsim 0.99$ produce occlusion maps which are qualitatively different from those shown in figure \ref{fig:Occ_SGLC2}; these maps tend to be relatively robust to the occlusion map process. There is a small number of mask locations that have a dramatic effect on the occlusion map that are either in the centre of the image or along the arcs of the lens. Masking the rest of the image induces a roughly equal suppression of the ``lens'' class output value.

\subsubsection*{Known Compound Lenses}
\begin{figure}
	\includegraphics[width=\columnwidth]{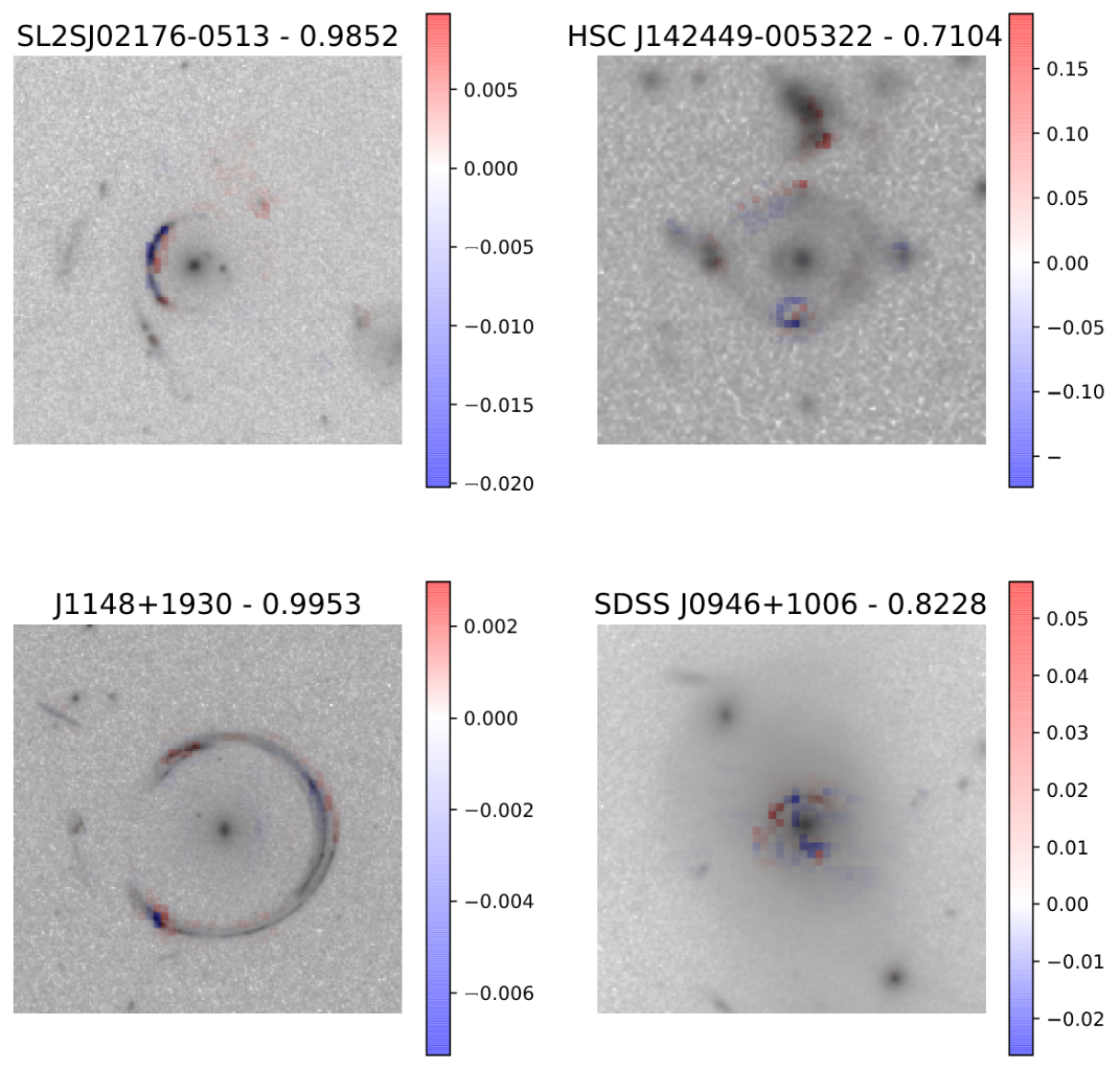}
    \caption{Occlusion Maps of known compound lenses. The left number shows the lens name for the image, and the right number is the model output for that image. The occlusion maps are independently scaled to highlight the change in output for each image, where the minimum value is blue, the maximum value is red, and the value of 0 is white. Blue indicates that the model associates the occulted feature with lensing and red indicates that the model associates the occulted feature with the non-lensing class. }
    \label{fig:CompoundLens_square_Occ}
\end{figure}

The occlusion maps for how OU-200 responds to the known compound lenses are shown in figure \ref{fig:CompoundLens_square_Occ}.

The occlusion map for SL2SJ02176-0513 highlights the arc with a strong lens response and the boundary of the arc has a strong non-lens response. 
The occulsion map for HSC J142449-005322 appears to be highlighting the quadruply lensed quasars in the image. This is unexpected behaviour as the training data only contains rings and arcs. The occlusion maps suggest that OU-200 may be identifying each quasar as an individual Einstein ring in the image, instead of identifying the two Einstein rings surrounding the central galaxy. In particular a clear ring-like feature is visible in the occlusion map close to the bottom quasar image. Another possibility is that no single $4 \times 4$ pixel square significantly affects the output of this image because of how large the Einstein rings are. Hence the model can compensate based on areas where the image is not occluded. 

The occlusion map for J1148+1930 highlights three main sections of the central Einstein ring. This is similar to the occlusion maps shown in \ref{fig:Occ_SGLC2}, where the model seems to detect the arcs in the image as lensing features. Again, the boundaries of the arcs seem to be detected as non-lens features. 

The occlusion map for SDSS J0946+1006 is difficult to diagnose. This is probably due to the spatial extension of the lensing features being relatively small compared to the size of the occlusion mask. It appears that the top arc is considered mainly as a non-lens feature and the bottom arc is mainly considered to be a lens feature. The fainter arcs of the large Einstein ring do not appear to be making any significant changes to the lens classification. 

\subsection{Images that activate the kernels}\label{sec:activeKernels}
Convolutional kernels are data structures associated with individual layers of a CNN that associate the trainable parameters (weights) of that layer with adjacent sets of pixels in its inputs.

For the first layer in the CNN, the kernel parameters map directly to adjacent sets of pixels in the input images. In later layers, each parameter maps to a larger region of input image pixels. 

As the CNN trains, kernels in the first layer learn representations simple shapes and colour relationships that are directly visible in the image pixels. Kernels in subsequent layers learn how to combine these into more abstract representations including, for example, adjacency and the global context of a simple feature within the image. Identifying which features the kernels have learnt over the course of training could help illuminate what features are important to the models' classification.

One of the ways to do this is to calculate the response of each kernel to each image. This is done by taking the mean value of the feature map generated by a kernel.

The mean value of a feature map (before apply ReLU activation) generated by a kernel is how the activation of a kernel is evaluated.
This is the kernel activation value (KAV). This is done to condense the output of a kernel into a single value. Images where the KAV is high, trigger a significant response in the kernel these are considered to highly activate that particular kernel. By comparing the images that have a high KAV for a given kernel, it can be inferred what a given kernel responds to. As to be expected with a CNN in the early layers of these models identify broad features of the image, such as the predominant colour of the image or if a large central galaxy is present. Looking deeper into the model more complex features start to emerge such as the position of a gravitational lens in the image or if the lensing feature is an arc or ring. 

Groups that could be identified from this approach include; large central galaxies, large Einstein rings, arcs, small blue Einstein rings, off centred Einstein rings, the position of large arcs, and the position of small blue Einstein rings. The large central galaxies that take up the majority of the image tend to be white, they are both spiral and elliptical, and the spiral galaxies can be either face or edge on. 

\begin{figure}
	\includegraphics[width=\columnwidth]{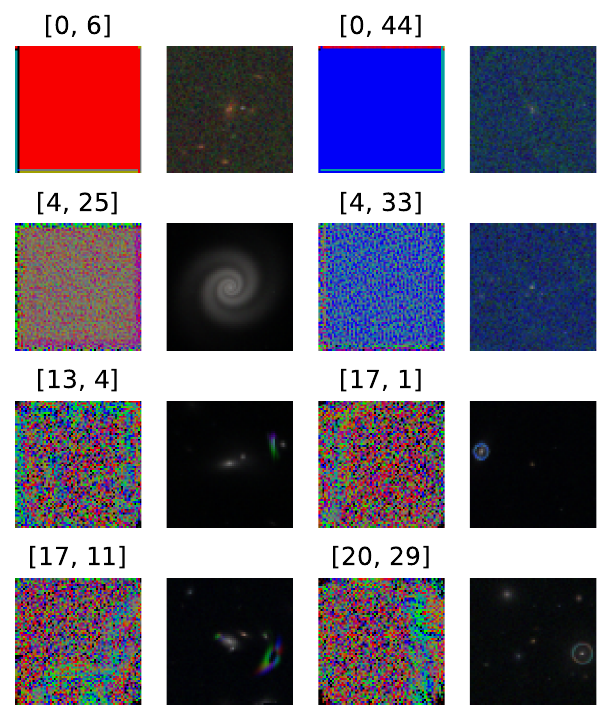}
    \caption{OU-66, \correctionsbold{First and third columns: Deep Dream creations that highly activate the kernel. Second and fouth Columns: The NISP image from the single lens dataset that most activates the kernel.} The square brackets indicate the layer number and kernel number for the kernel}
    \label{fig:TempMostActk66}
\end{figure}

\begin{figure}
	\includegraphics[width=\columnwidth]{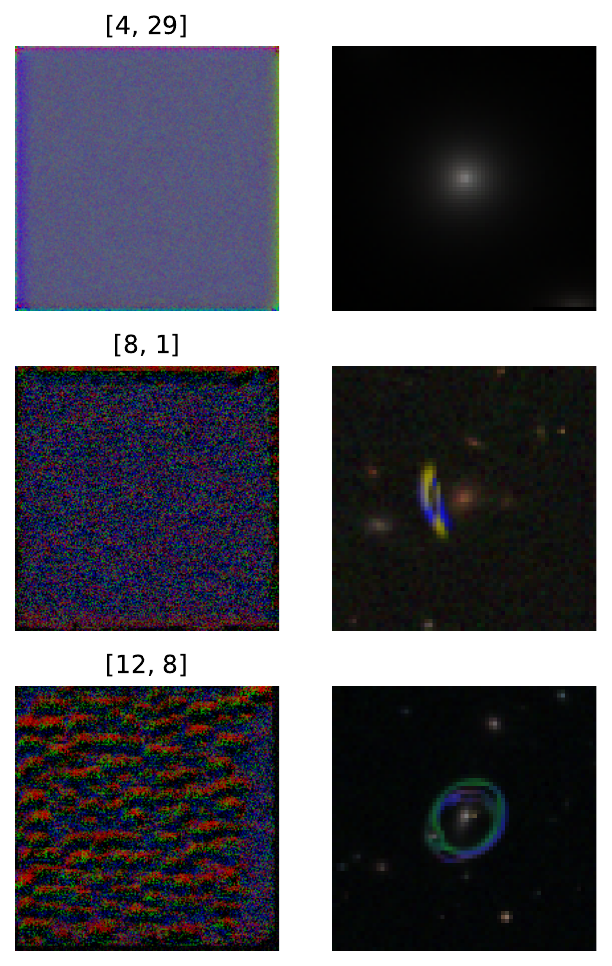}
    \caption{OU-200,  \correctionsbold{First and third columns: Deep Dream creations that highly activate the kernel. Second and fouth Columns: The NISP image from the single lens dataset that most activates the kernel.} The square brackets indicate the layer number and kernel number for the kernel}
    \label{fig:TempMostActk200}
\end{figure}

Figures \ref{fig:TempMostActk66} and \ref{fig:TempMostActk200} show in the right column the image which has the highest KAV. These figures show a range of the groups described above. In figure \ref{fig:TempMostActk66} [layer 0 kernel 6] shows an example of the group where the majority of galaxies within the image are red. [layer 0 kernel 44] and [layer 4 kernel 33] shows a group where the noise level is high in the blue bands. [layer 4 kernel 25] shows the group that consists of large spiral galaxies that take up the majority of the image. [layer 13 kernel 4] and [layer 20 kernel 29 shows the group where the lensing feature is in the right third of the image. [layer 17 kernel 1] shows the group where the lensing feature is in the left third of the image. [layer 17 kernel 11] is the group where the lensing feature is in the bottom right corner of the image. In figure \ref{fig:TempMostActk200}, [layer 4 kernel 29] shows the group which consists of a white pale blue central elliptical galaxy. [layer 8 kernel 1] shows the group where the galaxies in the image are red. [layer 12 kernel 18] shows the group which consists of central Einstein rings.

\subsection{Class Generated Images}\label{sec:classGenImages}
To understand what image features the model may be associating with lenses and non-lenses images were generated based upon the target class for the model \citep{yosinski2015understanding, simonyan2013deep}. The model weights were frozen and the output was maximised for the target class. Through back-propagation the input image was updated. This results in an updated image that activates the target class more strongly than the unmodified input image. This process was repeated to 
saturate the activation of the target class. This artificially generated new features in the image that cause a strong response for the target class. It is likely that these new features emulate those that the model has learnt to identify within the target classes. 

For OU-66, all 4 channels of the original input to this model are identical $66 \times 66$ pixel arrays containing random uniform noise on [0,1]. This process is repeated for OU-200 and the image is scaled up to $200 \times 200$ pixels.

Modified images were generated from random noise inputs using OU-66 and OU-200 for both classes. A learning rate of 0.1 was applied and the input image was updated by backpropagation 5000 times. The images produced at the end of this process can be seen on the top two rows of figure \ref{fig:DeepDream_All}. Both models produce similar outputs when generating a non-lens, creating a large blue and green blob in the centre of the image. This suggests that the model considers a non-lens to have large blue features in the centre of the image. The models generate different images for a lens. OU-66 creates a large blue Einstein ring that appears to be spatially offset in the 4 image bands. The area inside the ring is predominately red indicating that the model expects to see a red source galaxy in this region. OU-200 creates two separate structures, a small blob in the top left of the image and another diagonal feature going from bottom left to top right. The majority of the generated features are blue, but there are strong yellow and red curved features diagonally across the centre of the image. This could suggest that OU-66 is creating large central Einstein rings and that OU-200 is creating arc features within the image.

\subsection{Deep Dream}\label{sec:deepdream}
Deep dream is a technique that is similar to class generated images \citep{mordvintsev2015inceptionism}, but instead of the input being random data one of the simulated single lens images is used.  

\begin{figure}
	\includegraphics[width=\columnwidth]{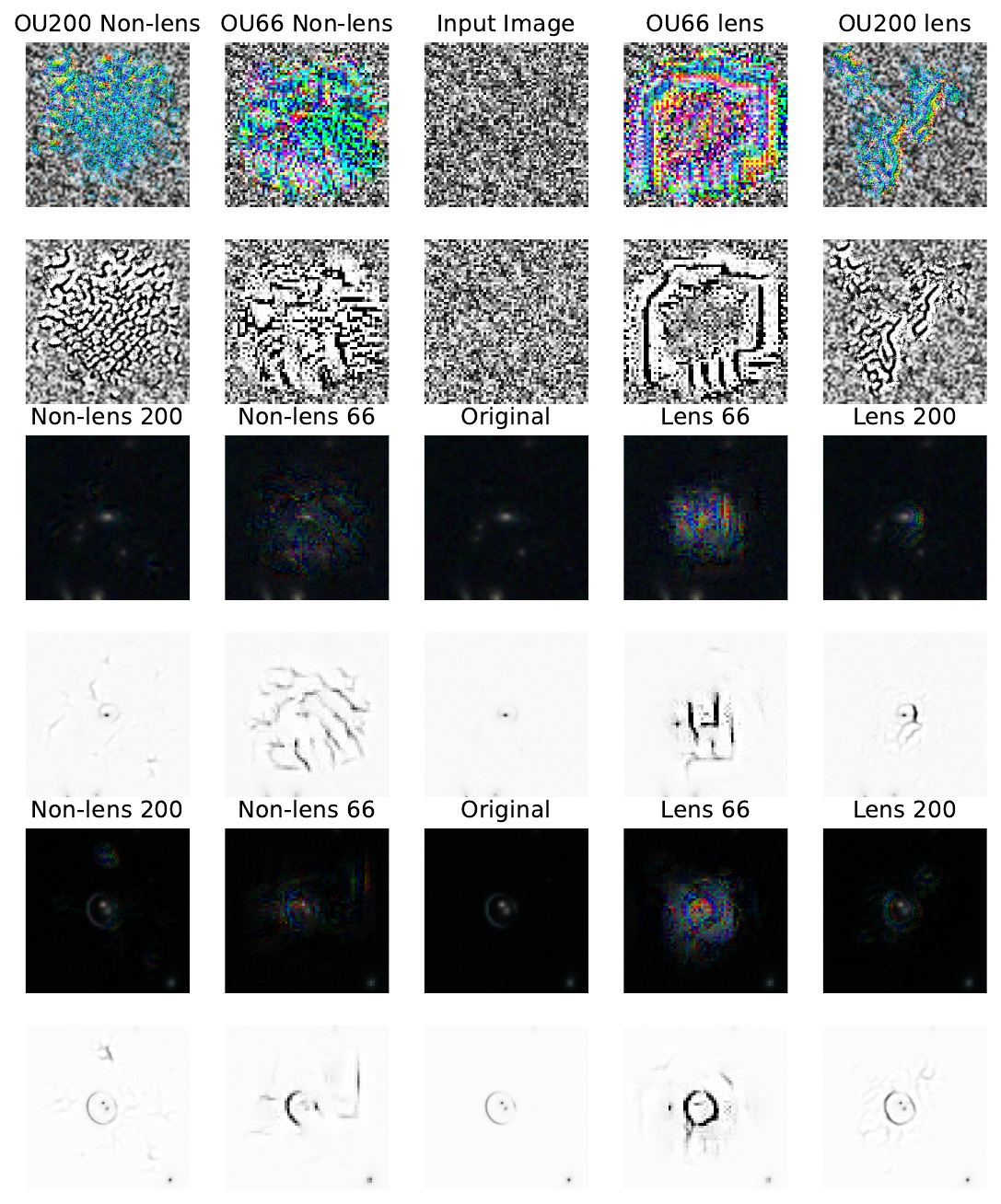}
    \caption{Rows 1 and 2: An example of a class generated images generated by both models for each target class. Rows 3 and 4: Deep Dream images created by both models for each target class starting from an image of a non-lens. Rows 5 and 6: Deep Dream images created by both models for each target class starting from an image of a lens. The NISP bands for the deep dream images have had their brightness increased by 150\% for clarity of the features generated.}
    \label{fig:DeepDream_All}
\end{figure}

The learning rate used was 0.001 and this process was allowed to run for 1,000 iterations. Examples of this technique being applied to a lens image and a non-lens image are shown in the bottom four rows in figure \ref{fig:DeepDream_All}. 

Rows three and four of figure \ref{fig:DeepDream_All} show Deep Dream images that result when the input image was a simulated non-lens, and ``non-lens'' was used as the target class. Deep Dream images created using OU-200 for ``non-lens'' images and target class differ very little from their inputs. In contrast, the corresponding Deep Dream images for OU-66 exhibit blue green diagonal features that could be representing galaxies. When ``lens'' is the target class both models created features around the central galaxy when using ``non-lens'' input images. OU-66 created a ``blocky'' arc below the central galaxy, while OU-200 created a rounder arc to the right of the central galaxy. Both models have a colour gradient across these generated features suggesting that the model is activated by a change in colour across features.

Rows five and six of figure \ref{fig:DeepDream_All} show Deep Dream images that were generated when the input image was a simulated single lens, and ``non-lens'' was used as the target class. In the OU-200 Deep Dream images the Einstein ring becomes red suggesting that this model heavily relies on colour for classification. Several round artefacts appear in the image, which might indicate that the model has learnt to associate multiple flux peaks in an image with the ``non-lens'' class. The Deep Dream image for OU-66 also fragments the Einstein ring into two sections and appears to straighten both sections. The resultant image begins to resemble edge-on spiral galaxies, indicating that the model may have learnt that the presence of such galaxies in images is also likely to indicate the ``non-lens'' class. This is reasonable since the majority of gravitational lenses in the simulated training data (and in reality) are large elliptical galaxies. Unlike OU-200, the Deep Dream image for OU-66 shows little evidence for colour modification which indicates that the model places less weight on colour and bases its classifications primarily on morphological information.

When ``lens'' was the target class both models make the Einstein ring brighter and bluer and tend to join previously incomplete arcs into complete Einstein rings. OU-66 often creates a red and yellow feature inside the Einstein ring. This feature could indicate that the model associates red central galaxies with images in the ``lens'' class. Again this is consistent with the prevalence of gravitational lenses that are the halos of red elliptical galaxies.

\subsection{Generating Images to Highly Activate Kernels}\label{sec:deepDreamKernels}

Although it is often used to produce images that activate a specific target class, the Deep Dream technique can also be used to generate images that activate a specific kernel within the model \citep[see e.g.][]{yosinski2015understanding, simonyan2013deep}. This will be referred to as Generated Kernel Images (GKI).

The central columns of Figures \ref{fig:TempMostActk66} and \ref{fig:TempMostActk200} show images that activate the kernels illustrated in the left-hand columns of those figures. The descriptions below outline some of the features these kernels detect and what they could relate to in the input images.

As can be seen in figure \ref{fig:TempMostActk66}, [layer 0 kernel 6] creates a solid red block with some other colours along the edge. [layer 0 kernel 44] creates a solid blue block with some other colours along the edge. [layer 4 kernel 25] creates a mixture of pink and green pixels.
[layer 4 kernel 33] creates a blue noise that appears very similar to the images which are highly activated by this corresponding kernel.
[layer 13 kernel 4], [layer 17 kernel 1], [layer 17 kernel 11], [layer 20 kernel 29], all of these deep dream images have the same overall feature, where the majority of the image is red apart from a section of the image which is very blue and green. The difference between these GKIs is the location and shape of these blue features. The position of these blue features corresponds to the location of the lens within the highly activating image.

In figure \ref{fig:TempMostActk200}, [layer 4 kernel 29] creates a pale blue feature with a low level of noise. [layer 8 kernel 1] creates an image with a black background with a mixture of noise that is red, green, and blue. [layer 12 kernel 8] is very similar to [layer 8 kernel 1], however this has several horizontal 'sand ripples' that are green and red which are slightly offset vertically from each other.  This feature could be trying to detect horizontal changes in colour.

\subsection{Guided Grad-CAM}\label{sec:guided_grad_cam}
Guided Grad-CAM \citep{selvaraju2017grad} is a common technique used for CNN interpretability. It is well-know for its association with the ImageNet challenge, where it has highlighted areas of images that correspond to areas that highly influence the determination of an output class in various models. The technique works by taking a weighted average of all the feature maps in the last convolutional layer with ReLU applied. This creates a Grad-CAM image highlightling locations in the image where the CNN detects features of interest. This Grad-CAM image only shows features that increase the model output value for the target class. Guided backpropagation \citep{springenberg2014striving} is applied to the image, which sets the desired class to 1 and all other outputs to 0. Only the positive gradients are backpropagated through the model (negative gradients are clipped at zero) to show areas of the image that have a positive effect on the classification. Once both processes are complete the results of guided backpropagation and Grad-CAM are multiplied together to generate a Guided Grad-CAM image. 

\begin{figure}

	\includegraphics[width=\columnwidth]{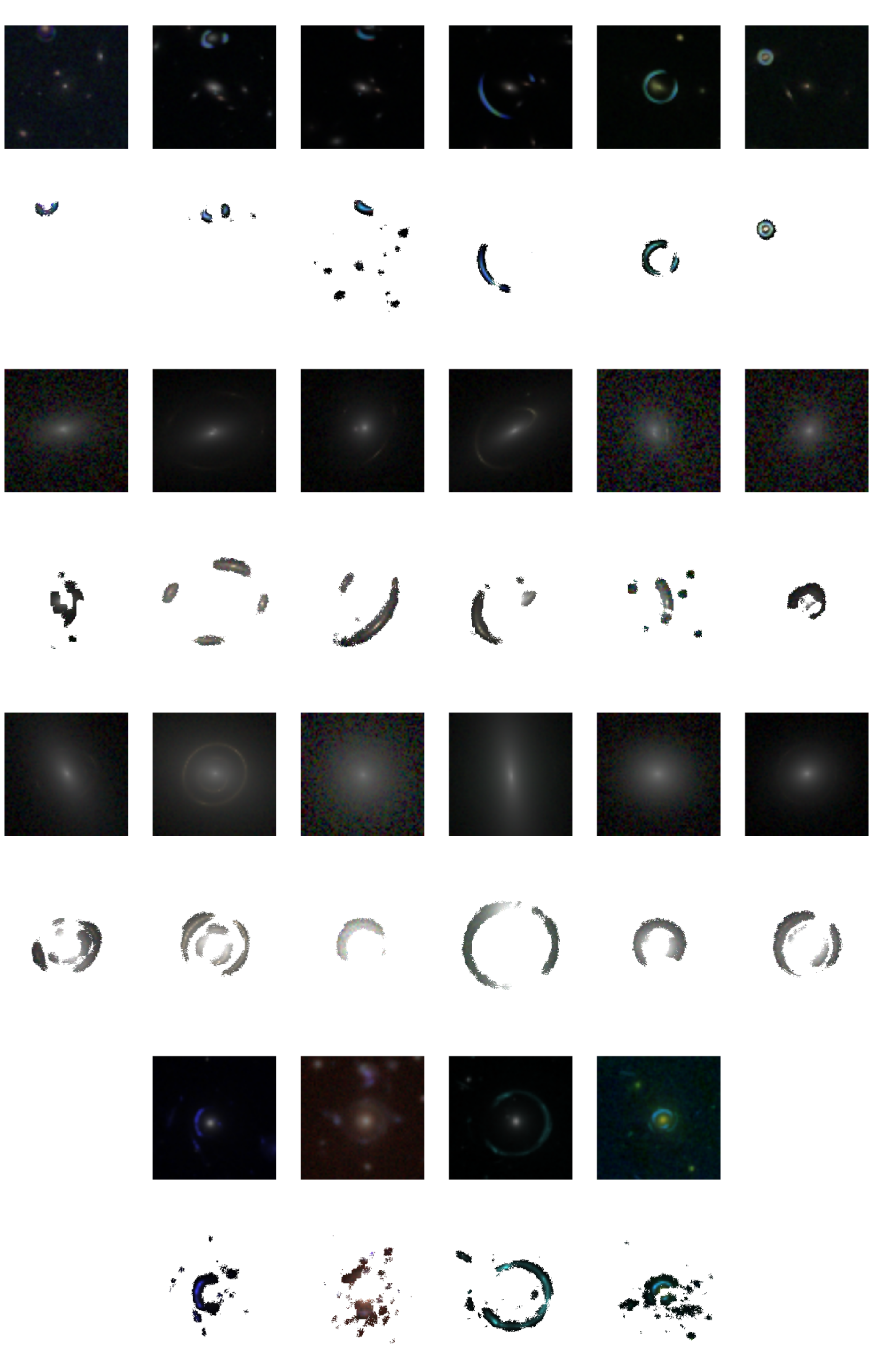}
    \caption{Top Row: \correctionsbold{Guided} Grad-CAM of single lens,
    Second Row: \correctionsbold{Guided} Grad-CAM of compound arcs,
    Third Row: \correctionsbold{Guided} Grad-CAM of double rings,
    Bottom Row: \correctionsbold{Guided} Grad-CAM of known compound lenses.}
    \label{fig:master_GradCAM}
\end{figure}

Guided Grad-CAM was applied to a set of images for the OU-200 model. In figure \ref{fig:master_GradCAM}, the input images are shown in rows 1, 3, 5 and 7. Rows 2, 4, 6, and 8 show the pixels from the input image when the pixel value in that position in the \correctionsbold{Guided} Grad-CAM image is greater than or equal to 0.05. The majority of these \correctionsbold{Guided} Grad-CAM images highlight the gravitational lens feature in the image. \correctionsbold{Guided} Grad-CAM finds parts of both rings in four of the six double ring images in figure \ref{fig:master_GradCAM}. For the known compound lenses, \correctionsbold{Guided} Grad-CAM selects large sections of the gravitational lensing features for SL2S J02176-0513 and SDSS J1148+1930. For SDSS J0946+1006, \correctionsbold{Guided} Grad-CAM mainly selects the brightest lensing arc and for SL2SJ02176-0513 small sections of the Einstein ring are selected. In most of the known compound lens images \correctionsbold{Guided} Grad-CAM does not find features from both lenses in the image, with the exception of SDSS J0946+1006 where a small area of the fainter lens is clearly highlighted.

\section{Discussion}\label{sec:discussion}
In this paper, several Deep Learning models have been applied to classify  simulated single lenses and non-lenses, simulated compound lenses, and previously discovered compound lenses. Models that process multiple bands (in particular those that include the bluer VIS band) with higher angular resolution like OU-200 tend to improve the recall for the gravitational lens class. Providing high-resolution multi-band input images provides the model with additional information during training, which allows it to make more accurate predictions when determining whether an image contains a lens. OU-200 was shown to adapt well to a simulated compound lens dataset that the model had not been trained on. This indicates that training models to accurately identify single lens systems does not necessarily preclude their use for identifying compound lens systems. Multi-band models (OU-66 and OU-200) are able to classify all the previously discovered compound lenses shown in this paper as lenses. These models can adapt from simulated images to observational data including HST and HSC images, suggesting that they will need little retraining to perform well on real images in future upcoming surveys. This has previously been shown by the CNN described in \citep{petrillo2019testing} which was trained on KiDS data and performed second best in the strong gravitational lens challenge \citep{metcalf2019strong} without retraining, demonstrating that these CNNs can find gravitational lenses outside of the data they have been trained on. 

In section \ref{sec:interpretability}, the factors affecting the classifications made by the models in this paper were investigated. Features associated with lensing such as Einstein rings and arcs, significantly influence the output of these models, which implies that they have correctly learnt to associate these features with the lens class. Based on the Deep Dream images that were generated to activate specific kernels ($\S$\ref{sec:deepDreamKernels}), the class generated images ($\S$\ref{sec:classGenImages}), and the deep dream images ($\S$\ref{sec:deepdream}) it is clear that colour is important to the models' classifications of the images. Inspecting the GKIs revealed kernels in the deep layers of the models are strongly activated by blue regions on red backgrounds at the location of the gravitational lens. For OU-66, the class generated images ($\S$\ref{sec:classGenImages}) exhibit blue ring-like structures enclosing a red central region. This indicates that this model has learnt that blue ring-like-structures surrounding central red features, which are typical of many gravitational lensing systems, are more likely to indicate the image contains a gravitational lens. Pessimistically, this might indicate that this model would struggle to classify images with red source galaxies as containing a gravitational lens and more generally that the model may have a colour bias. This colour bias has also been reported in \citep{petrillo2019testing, jacobs2022exploring}. This is not necessarily a bad thing for the model to have learnt, since the majority of known gravitational lens systems have blue source galaxies and red lensing galaxies. Nevertheless, it selects against systems that are not typical of currently known lenses.

\correctionsbold{The only way to begin to understand the biases within methods such as CNNs for gravitational lens detection is to investigate the interpretability of these methods. This is possible through a variety of means such as investigating the parameters learnt by these models directly as shown in this paper or by manipulating the input images via a sensitivity probe \citep{jacobs2022exploring}. The use of a sensitivity probe has been able to show how susceptible CNNs can be to minor modifications to the input data. The indications of small decisions significantly affecting the performance of CNNs are present. Features such as the PSF can significantly affect the ability of a CNN to learn data (as we discuss later on) or the ability of a CNN to adapt to new data. }

Accordingly, OU-200 has learnt relevant details about typical gravitational lenses, but this also shows that this model may not adapt well when presented with a gravitational lens system where the source galaxy is not blue. The class-targeted deep dream images ($\S$\ref{sec:deepdream}) also exhibit blue, ring-like features surrounding red central regions. Moreover, when targeting the ``lens'' class with a lens image as input, the lens is emphasised and becomes bluer. When targeting the non-lens class using a lens image as input the Einstein ring becomes fragmented in OU-66 and redder in OU-200, indicating that the shapes of features are more important to OU-66 when classifying images, while instead, colour information is more strongly weighted by OU-200. OU-200 has been shown in figure \ref{fig:DeepDream_All} to actively associate the ``non-lens'' class with red Einstein rings. Whereas the class generated images of the ``lens'' class generate blue Einstein rings, suggesting that the model may not classify images with red Einstein rings as lenses.

This hints at a fundamental limitation to the current approaches to lens finding with missions and surveys such as {\it Euclid} and LSST: the training sets used for the algorithms may inadvertently select against non-negligible populations of strong gravitational lenses that exist on the sky. 
There are now several known examples of gravitational lensing systems that exhibit red Einstein rings \citep{geach2015}, and there are growing numbers of known gravitationally lensed dusty star-forming galaxies \citep[e.g.][]{negrello2010,negrello2017,spt2016,act2019,neri2020,urquhart2022bright}. 
The surface density of submm-bright gravitational lenses (e.g. $500$\microns\ flux densities $S_{500}\geq100$\,mJy) is relatively low, with only one lens per 7.5\,deg$^2$ \citep[e.g.][]{negrello2017}, compared to e.g. $\sim$50 lenses per square degree in COSMOS \citep[e.g.][]{faure2008,jackson2008}, but these are only the lensed submm galaxies that are easy to identify in blank-field surveys. At submm flux densities $\sim10\times$ fainter, the predicted surface density of strong lenses is expected to be about two orders of magnitude higher, e.g. $\sim10$ strong lenses per deg$^2$ with $S_{500}\geq10$\,mJy \citep[e.g.][]{trombetti2021}. This implies $\sim150\,000$  strong submm-wave lenses over the entire {\it Euclid} survey area, comparable to the total expected to be found with the {\it Euclid} VIS instrument \citep[e.g.][]{collett2015}. As one probes fainter submm luminosities and lower dust obscurations, it is reasonable to expect the background sources to become increasingly detectable at near-infrared wavelengths, suggesting the possibility of a non-negligible exotic population of lensed red galaxies in wide-field surveys such as {\it Euclid}. Furthermore, extreme ultra-high-redshift lensed galaxies will be preferentially detected only at near-infrared wavelengths, and though very rare on the sky \citep[e.g.][]{marchetti2017,vikaeus2021}, will be high priority targets for follow-ups. Our results show that it is not yet clear whether such red background lensed sources will be systematically selected against with default lens finders.

On the other hand, the sensitivity to morphological features has proven promising for the discovery of compound lens systems.
The \correctionsbold{Guided} Grad-CAM images shown in figure \ref{fig:master_GradCAM}, highlight the lensing features in the images for the majority of the examples shown. This indicates that the model has learnt to look for indicative features in the image such as rings and arcs. The CNNs presented in this paper appear to be sufficiently flexible that they correctly classify exotic systems with multiple lensing planes, despite no such systems appearing in the datasets on which they were trained. This flexibility is encouraging because it suggests that similar exotic systems that exist in the real Universe will not be missed when the models are applied to real upcoming observational data. Nevertheless, while the successes in detecting compound lenses are encouraging, it is likely that improvements could be made by training the networks explicitly to recover them.

When initially training the CNNs in this paper there was a class imbalance in the training data. To balance this dataset the non-lenses were rotated by $90\deg$, $180\deg$, $270\deg$ and flipped along the horizontial axis then rotated by $90\deg$, $180\deg$, and $270\deg$. This caused the model to classify images which had not been augmented as lenses, and any image which had been augmented was classified as a non-lens. The model had learnt to classify the images based upon their orientation. It was realised that the model could determine the orientation image based on the orientation of the simulated Euclid PSF, i.e. its three-fold symmetry meant the model could separate images based on the augmentations that had been applied to it. This highlights the sensitivity of machine learning techniques like CNNs to very subtle, and scientifically irrelevant, image features that would be completely imperceptible for human inspectors. We would like the network outputs to be invariant with respect to the astrophysical orientation on the sky, but the detector orientation imposes rotational asymmetry by virtue of the PSF shape. One possible solution could be to add the detector orientation on the sky as an auxiliary input to the network training and the prediction. This would reintroduce the ability to learn rotational invariance from the training data by augmenting them using rotation, or allow the application of group-equivariant CNNs \citep{scaife2021fanaroff}.

The application of machine learning to lens finding in this paper has been predicated on the intractability of exhaustive human inspection, but even an apparently-perfect performance on simulated data need not imply that there is no useful contribution to be made by human volunteers \citep[e.g.][]{marshall_spacewarps_2016,more2016space,sonnenfeld_spacewarps_2020}. The sample of candidate gravitational lenses is likely to be of a tractable size for volunteers or experts to examine every one. This may identify under-performance or other problems for known rare sub-populations, which may in turn be used to refine or retrain automated lens finders, in a virtuous circle between human and machine vision. Human inspection is also currently the only widely-applied method for finding new, unanticipated categories of exotic systems. Machine learning approaches are in development to detect anomalies in astronomical imaging, such as training a generative adversarial network (GAN) to encode the space of commonly-occurring features in real astrophysical image \citep[e.g.][]{storey-fisher2021}. Atypical images can then be identified as those for which the discriminator of the GAN rejects the image. This has succeeded in identifying images affected by instrumental defects and/or artefacts, but there are still formidable obstacles in representing the diversity of (for example) galaxy morphologies, and ``interesting'' anomalies are still a matter of human judgement.

\section{Conclusions}\label{sec:conclusions}
One of the main advantages of very large samples of strong gravitational lenses is the potential for discovery of rare lens configurations, among which the compound lens systems have the added bonus of providing new cosmological parameter constraints \citep[e.g.][]{gavazzi2008sloan}. This places the detection of these systems at a premium within survey projects such as {\it Euclid} and LSST. Our analysis has shown that existing lens finding methodologies should be able to recover compound lensing systems even without further training, albeit not with 100\% recall. Ultra-high-redshift lensed sources will be at a similar high premium, but it is less clear that these red lensed systems will be detectable by the lens finder presented in this paper. It seems clear that, in order to be able to exploit the advantages of a very large strong lensing sample, attention must be paid to training the lens finding algorithms to accommodate the anticipated interesting subsets. For the ``unknown unknowns'', however, the data mining problem remains intractable by definition, and there may be no option other than a coordinated programme of human inspection such as through citizen science \citep[e.g.][]{marshall_spacewarps_2016,more2016space,sonnenfeld_spacewarps_2020}.

\section*{Acknowledgements}
%The Acknowledgements section is not numbered. Here you can thank helpful
%colleagues, acknowledge funding agencies, telescopes and facilities used etc.
%Try to keep it short.

JW was supported by the Science and Technology Facilities Council under grant number ST/P006760/1,  the DISCnet Centre for Doctoral Training in Data-Intensive Science.
SS and HD were partly supported by the ESCAPE project; ESCAPE - The European Science Cluster of Astronomy \& Particle Physics ESFRI Research Infrastructures has received funding from the European Union's Horizon 2020 research and innovation programme under Grant Agreement no. 824064. SS also thanks the Science and Technology Facilities Council for financial support under grant ST/P000584/1.

\section*{Data Availability}

The code used in this paper is available at \url{https://github.com/JoshWilde/LensFindery-McLensFinderFace} \github{https://github.com/JoshWilde/LensFindery-McLensFinderFace}\,.

%%%%%%%%%%%%%%%%%%%%%%%%%%%%%%%%%%%%%%%%%%%%%%%%%%

%%%%%%%%%%%%%%%%%%%% REFERENCES %%%%%%%%%%%%%%%%%%

% The best way to enter references is to use BibTeX:

\bibliographystyle{mnras}
%\bibliography{example} % if your bibtex file is called example.bib
\bibliography{bibliography} % if your bibtex file is called example.bib

% Alternatively you could enter them by hand, like this:
% This method is tedious and prone to error if you have lots of references
%\begin{thebibliography}{99}
%\bibitem[\protect\citeauthoryear{Author}{2012}]{Author2012}
%Author A.~N., 2013, Journal of Improbable Astronomy, 1, 1
%\bibitem[\protect\citeauthoryear{Others}{2013}]{Others2013}
%Others S., 2012, Journal of Interesting Stuff, 17, 198
%\end{thebibliography}

%%%%%%%%%%%%%%%%%%%%%%%%%%%%%%%%%%%%%%%%%%%%%%%%%%

%%%%%%%%%%%%%%%%% APPENDICES %%%%%%%%%%%%%%%%%%%%%

\appendix

%%%%%%%%%%%%%%%%%%%%%%%%%%%%%%%%%%%%%%%%%%%%%%%%%%

% Don't change these lines
\bsp	% typesetting comment
\label{lastpage}
\end{document}